# Dialectics of antimicrobial peptides II: Theoretical models of pore formation and membrane protection

Oleg V. Kondrashov[1], Marta V. Volovik[1], Zaret G. Denieva[1], Polina K. Gifer[1], Timur R. Galimzyanov[1, #], Peter I. Kuzmin[1], Oleg V. Batishchev[1], Sergey A. Akimov[1]

[1]Laboratory of Bioelectrochemistry, A.N. Frumkin Institute of Physical Chemistry and Electrochemistry, Russian Academy of Sciences, 31/4 Leninskiy prospekt, Moscow, 119071, Russia

[#]Present address: JetBrains, JetBrains Research, 23 Christoph-Rapparini-Bogen, München, 80639, Germany

Correspondence: olegbati@gmail.com (O.V.B.); akimov_sergey@mail.ru (S.A.A.)

Amphipathic peptides are considered promising antibiotics because of their ability to form pores in bacterial membranes. In two companion papers, we analyzed both experimentally and theoretically the mechanisms and consequences of the interaction of two types of amphipathic peptides (magainin and melittin) with lipid membranes. We studied this interaction for different peptide concentration: low, high, and low concentration followed by the addition of peptides in high concentration. Here we provide the theoretical description of the pore formation mechanisms. We predicted theoretically that two peptide molecules are enough to locally induce the formation of a small metastable pore that continuously connects two membrane leaflets and allows peptide and lipid translocation between the leaflets. This mechanism (referred to as local) is supposed to work at low peptide concentrations. When applied in high concentration, the one-sided adsorption of peptides onto a closed membrane generates lateral pressure in the contacting lipid monolayer and lateral tension in the opposing monolayer. Our calculations predicted such asymmetric pressure/tension to greatly facilitate the formation of large metastable pores at any point of the membrane, regardless of the distance to the nearest peptide molecule. We therefore refer to this mechanism of pore formation as non-local. When the application of peptides in low concentration is followed by high concentration addition, multiple small metastable pores are predicted to form in the membrane in accordance with the local mechanism. This prevents the generation of a large difference in lateral pressure/tension, thus protecting the membrane from the formation of large pores. The results of the theoretical analysis agree with the experimental data of the companion paper.

# I. INTRODUCTION

Various amphipathic peptides are known to form pores in bilayer lipid membrane [1–4]. Due to this activity, some $\alpha$-helical and $\beta$-hairpin peptides are considered as promising antimicrobial agents with a predicted low probability of drug resistance [1,2]. A negative electric charge is exposed at the outer monolayer of bacterial membranes, while the outer surface of plasma membranes of eukaryotic cells is electrically neutral. The selectivity of antimicrobial peptides (AMPs) is mainly based on this difference, and most AMPs carry high positive electric charge. E.g., the electric charge of melittin from the bee venom is $+6e$ [5–7], of magainin from the venom of the African frog *Xenopus laevis* is $+5e$ [5,8], of crotalicidin from the venom of South American rattlesnake is $+16e$ [1]. Two general types of pores are formed by AMPs: 1) barrel-stave pores, where the pore edge is completely lined with peptide molecules; 2) toroidal pores, where the edge is formed by both peptide and lipid molecules [9–12]. As a rule, barrel-stave pores have a fixed stoichiometry; they are formed by assemblies of β-hairpin peptides, or $\alpha$-helical peptides having a large hydrophobic side-surface area [10,13]. The stoichiometry of toroidal pores is not fixed. These pores are usually formed by $\alpha$-helical peptides having a relatively small fraction of the hydrophobic area of the helix side-surface [14]. Amphipathic peptides are thought to exist in membranes in two long-living configurations: 1) S-configuration, in which a longitudinal axis of the helix is nearly parallel to the membrane plane; 2) I-configuration, in which there is a substantial angle between the longitudinal axis of the helix and the membrane plane [15,16]. In both configurations, peptide molecules are partially inserted into the lipid monolayer of the membrane, which is detected by the increase in the total membrane area [4,17,18]. The I-configuration corresponds to a through barrel-stave or toroidal pore formed in the membrane [10,11]. A molecular mechanism and even physical reasons for the transition from S- to I-configurations and vice versa are not actually known. A direct immersing of highly charged AMP molecule(s) into the hydrophobic interior of a membrane seems quite inconceivable. Most probably, the transition from S- to I-configuration should proceed via some intermediate states, implying a gradual re-orientation of AMP molecules and rearrangement of surrounding lipids. However, these hypothetical intermediate states are too short-living to be observed experimentally. Even more counterintuitive is the existence of amphipathic cell-penetrating peptides (CPP). These peptides are highly charged (e.g., $+7e$) [19], and when added to the water solution at one side of the membrane, after some time are found in the water solution at the opposite side. In addition, they are able to penetrate multilamellar vesicles [20].

The penetration through membranes usually occurs without the formation of macroscopic pores, although some authors report a peptide-induced leakage of the membrane with respect to fluorescent molecules like calcein [20,21].

Amphipathic peptides partially inserted into a lipid monolayer unavoidably deform the membrane. The characteristic length of the deformation decay in a typical membrane is about several nanometers [22,23]. When peptide molecules are far apart, the induced deformations are independent and their energy is additive. Upon mutual approach, the deformations overlap, leading to an effective lateral interaction of peptides. Pore formation becomes intense when the AMP surface concentration on the membrane exceeds some critical value [1,4,15]. This may indicate that pore formation by AMPs is a cooperative process. However, the critical concentration weakly depends on the detailed chemical structure of the peptide, and for most AMPs lies in the range of 1/25–1/100 peptide/lipid [1,4,15]. At these surface concentrations, the membrane is almost completely covered by deformations induced by inserted AMP molecules [22]. Therefore one can infer that the critical concentration is not associated with the formation of any cooperative structure of a certain stoichiometry, since its formation should strongly depend on the specific primary/secondary structure of the peptide. In addition, tight interaction and assembly of a highly charged multimer are very doubtful, as the charged molecules should strongly repel each other at distances < 1 nm, which is the Debye screening length under physiological conditions.

One of the earliest models of pore formation in a thin structureless film has been developed by Derjaguin and Gutop [24]. The free energy of formation of a circular pore of radius $R$ in the film is given by: $E = 2\pi R\gamma - \pi R^2\sigma$, where $\gamma$ is the line tension of the pore edge; $\sigma$ is the lateral tension of the film. The two terms correspond to: 1) an excessive energy of the film material at the pore edge as compared to its energy far from the pore (internal energy); 2) work of external forces (lateral tension) in the process of pore formation. The dependence $E(R)$ has a maximum at the critical radius $R^* = \gamma/\sigma$. In order to form a supercritical pore ($R > R^*$), an energy barrier of height $\Delta E = E(R^*) = \pi\gamma^2/\sigma$ should be surmounted. The Derjaguin-Gutop model is qualitative, as lipid membranes are neither infinitely thin nor structureless. Recent work on the structure and energy of the pore edge has shown that the line tension depends on both the radius of the pore and the lateral tension of the membrane, i.e., $\gamma = \gamma(R, \sigma)$ [25]. Nevertheless, the Derjaguin-Gutop model provides qualitative insights into the possible mechanism of AMP pore-forming activity. In typical lipid membranes pores rarely form spontaneously, meaning that the energy barrier of pore formation should be high. The height of the energy barrier, $\Delta E = \pi\gamma^2/\sigma$, can be reduced by decreasing the line tension or by increasing the lateral tension.

The structure of the pore of zero radius differs from the intact planar bilayer [Fig. 1(a)]. Lipid molecules located in the vicinity of the plane of symmetry are oriented horizontally [Fig. 1(b)], while in the horizontal intact bilayer all lipid molecules are vertical [Fig. 1(a)]. This means that the process of pore formation should pass through some intermediate state(s) with a gradual change of the orientation of the lipid molecules from vertical to the horizontal one. The so-called hydrophobic defect has been proposed as such an intermediate state [26–28]. In this state, lipid molecules are displaced radially from some axis, which is perpendicular to the membrane plane, and form a cylinder filled with water. The side walls of the cylinder are lined with hydrophobic lipid tails [Fig. 1(c)]. The length $L$ of the hydrophobic cylinder represents an additional degree of freedom of the system, meaning that two coordinates are needed to unambiguously define the state of the membrane with the pore: $R_0$ and $L$ [25]. In the process of pore formation via the hydrophobic defect the coordinate $L$ monotonically decreases from $2h$ to 0, where $h$ is the thickness of the hydrophobic part of the unperturbed lipid monolayer [25].

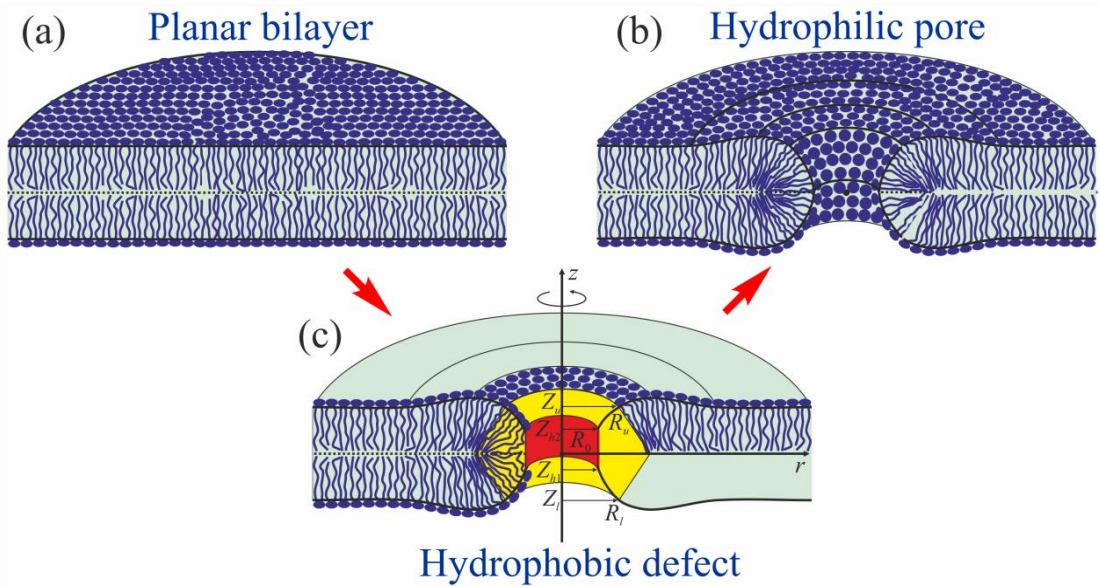

FIG. 1. Trajectory of pore formation from planar lipid bilayer (a) to hydrophilic pore (b) via hydrophobic defect (c). The hydrophobic cylinder is shown in red; the vertical monolayers are shown in yellow. The radius of the hydrophobic defect is $R_0$; the length of the hydrophobic cylinder is $L = Z_{h2} - Z_{h1}$. The vertical monolayer meets the upper horizontal monolayer along the circle $\{R_u, Z_u\}$; the lower monolayer — along the circle $\{R_l, Z_l\}$ in cylindrical coordinates.

Experiments on single giant unilamellar vesicles (GUVs) attached to a glass micropipette provide information on the interaction of AMPs with lipid membranes [4,17,18]. When one adds AMP to the GUV exterior, this leads to its adsorption and partial insertion into the outer leaflet of the GUV membrane. As GUV membrane comprises a finite number of lipid molecules, i.e., it lacks a lipid reservoir, such an insertion should lead to an increase in the area of the outer leaflet. However, the outer and inner leaflets have a common monolayer interface, and no voids are supposed between the leaflets, i.e., the leaflets have approximately equal areas. Thus, the increase in the area of the outer leaflet should be accompanied by about the same increase in the area of the inner leaflet, resulting in a stretching of the inner leaflet [29,30]. A simple consideration based on Hooke's law predicts that when amphipathic peptides, inserted into the outer leaflet of the membrane add a lateral area $S_p$, the outer leaflet should laterally compress by the area $S_p/2$, while the inner leaflet should laterally stretch by the area $S_p/2$; the overall increase in the membrane area upon the peptide insertion should be $S_p/2$. This leads to the generation of lateral pressure in the outer leaflet and lateral tension in the inner leaflet of the membrane. If the bulk concentration of AMP is high enough, after a stochastic lag time the GUV would abruptly draw completely into the micropipette, indicating the formation of a macroscopic pore(s) able to transmit a large amount of water from the GUV interior [4,17,18]. In many works, the asymmetric lateral pressure in the outer leaflet and the lateral tension in the inner leaflet of the GUV membrane generated by the adsorbing AMPs are considered the main reason for the macroscopic pore formation [29,30]. Both pressure and tension induce an elastic stress in the membrane, which can be partially released upon the formation of a pore. According to the Derjaguin-Gutop theory, the tension should facilitate pore formation, while the influence of the lateral pressure is not so obvious. Lateral pressure/tension are non-local features inherent to the whole membrane, independently on the distance to the closest AMP molecule.

An alternative way in which AMP can promote pore formation is based on a partial substitution of the lipid material at the highly deformed pore edge by a rigid, undeformable AMP molecule. This should lead to a decrease in the edge specific energy, i.e., in the line tension. In this case AMP promotes pore formation locally, in its immediate vicinity. In experiments on the interaction of PGLa with single GUV [4], the authors detect two equal steps of membrane area increase by 3% upon addition of the peptide in the concentration of 1.2 μM and 2.9 μM. They interpret this result as a formation of small hydrophilic pre-pores in the GUV membrane allowing the translocation of peptides into the GUV. In experiments with fluorescently labeled CPP adsorbing on GUVs loaded with a water-soluble dye, the membrane fluorescence is observed first. Then, this fluorescence intensity doubles, which can be interpreted as CPP translocation to the inner membrane leaflet, while vacant places at the outer leaflet become occupied by newly adsorbed CPP molecules. Upon an increase in the CPP concentration in the GUV environment, the leakage of water-soluble dye starts after the doubling of the membrane fluorescence [20,21]. From these data one can infer that AMPs form at least two types of pores: 1) Small pores, the formation of which does not lead to the irreversible rupture of the GUV membrane; 2) Large pores, leading to the irreversible rupture of GUV membrane. The formation of small pores or defects does not lead to an irreversible rupture of the GUV membrane subjected to a low background lateral tension (~0.5 mN/m) [4,17,18]. These defects arise at a low bulk concentration of AMP. The size of small pores allows passing PGLa molecule via their edge, while the pore lumen is small enough to prevent a fast water leakage from the GUV interior and complete suction of the GUV into the micropipette. In addition, the size of small pores (at least, formed by CPPs) is not sufficient for passing (leakage) of water-soluble dyes like Alexa Fluor 647 (AF647) [20]. Large pores formed by AMP at high bulk concentrations lead to a complete suction of the GUV membrane into the micropipette and to leakage of water-soluble fluorescent dyes like calcein or AF647 [4]. Such pores are most likely formed when many AMP molecules incorporate fast and simultaneously into the outer leaflet of the GUV membrane, leading to irregular large gradients of lateral pressure/tension in the membrane. According to the qualitative Derjaguin-Gutop model, in order to form small pores in a low concentration conditions, AMP should directly decrease the edge line tension. This can only be achieved if such small pores are formed in the immediate vicinity of the AMP molecule(s) or if AMP molecules directly participate in the formation of the pore edge. Thus, generally, AMP can perforate a membrane by two fundamental mechanisms: 1) non-locally, by modifying lateral pressure/tension [Fig. 2(a)]; 2) locally, by the direct participation in the formation of the pore edge [Fig. 2(b)].

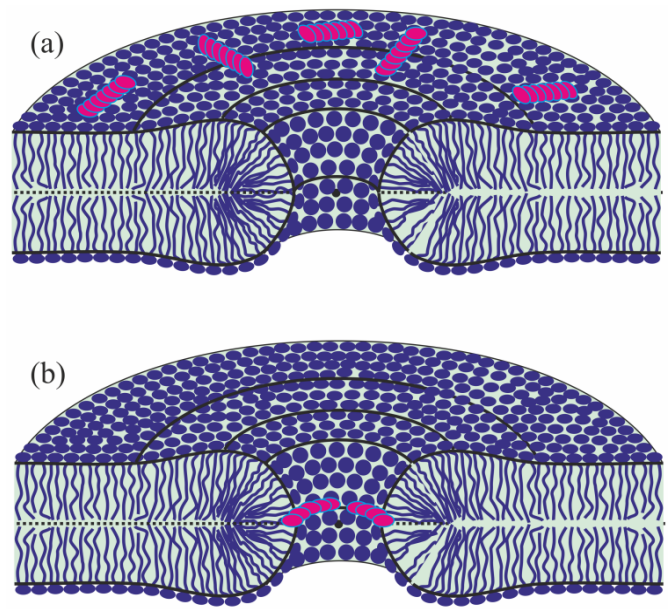

FIG. 2. Schematic illustration of non-local (a) and local (b) mechanisms of pore formation by amphipathic peptides. Peptides are shown in red color, lipids — in dark blue. According to the non-local mechanism (a), peptides adsorbed to one lipid monolayer generate lateral pressure in this monolayer and lateral tension in the opposing monolayer; this leads to formation of lipidic pores at any point of the membrane. In the local mechanism (b) peptide molecules directly participate in the formation of the pore edge thus forming lipid-peptide pores.

According to the Derjaguin-Gutop model, the energy barrier of pore formation is determined by the line tension of the pore edge and the lateral tension of the membrane. The barrier decreases for lower line tension and higher lateral tension. When the pore is formed by the local mechanism, peptide molecules directly participate in the formation of the pore edge [Fig. 2(b)], thus decreasing the line tension. When the pore is formed by the non-local mechanism, adsorbed peptides modify the lateral tension of the whole membrane [Fig. 2(a)]. Both mechanisms reduce the energy barrier of pore formation and produce metastable pores with a finite lifetime.

The membrane activity of various AMPs has been studied on large unilamellar vesicles (LUV). Peptide adsorption and desorption, as well as pore formation can be observed by means of fluorescence spectroscopy in such model system. The tryptophan amino acid residue is fluorescent which allows fluorescent imaging of the tryptophan-containing AMP. LUVs can be loaded with water-soluble fluorescent dye and/or fluorescence quencher. The LUV membrane can be labeled by fluorescent analogues of lipids. By combining various approaches (observation of direct fluorescence intensity, resonance energy transfer, etc.) and using various dyes and quenchers, it is possible to establish detailed kinetic schemes of the AMP activity. In a series of works [31–34] reviewed in [35], the kinetic schemes were proposed and experimentally proved for several AMPs. It is shown that amphipathic peptides can make LUV membrane permeable by two mechanisms: all-or-none (either complete or no leakage of the water-soluble dye from LUVs) and graded (continuous distribution of the concentration of the water-soluble dye in LUVs). However, these mechanisms seem to characterize the experimental system used to a greater extent than the size and stability of the AMP-induced pores. The definition of the mechanisms depends on the volume of the dye-loaded vesicles and the concentration of AMP. Indeed, in the limit of infinitely large giant unilamellar vesicles (e.g., 300 μm diameter) and low concentration of AMP, the probability of observation of the water-soluble dye distribution predicted by the all-or-none mechanism becomes zero: within the typical time of the experiment, all vesicles would lose only a fraction of the dye, which corresponds to the graded mechanism of the AMP activity. On the contrary, in the limit of very small vesicles (e.g., 30 nm diameter) and high AMP concentration, the probability of observing the graded release of the water-soluble dye becomes zero. These two types of experiments can be performed for the same AMP and membranes of the same composition, thus yielding that the particular mechanism of AMP activity (graded or all-or-none) should strongly depend on the particular experimental conditions, e.g., on the vesicle sizes, solution viscosity, dye molecule size, lifetime of the pore, etc. AMPs cannot be simply attributed to release the dye by only all-or-none or graded mechanism. The mechanism is not the characteristics of the peptide itself, it is rather the characteristics of experimental conditions.

In a number of papers [31–34] reviewed in [35], the authors have quantitatively determined the rate constants of AMP adsorption and desorption, as well as pore state formation, within the proposed kinetic schemes. The results obtained allowed to propose mechanisms of AMP activity and to qualitatively predict the structure of AMP-induced pores. In the conclusion of work [31], the authors write that for AMP cecropin A: "The molecular structure of the pore state still needs to be determined, but does not seem to involve any specific peptide arrangement."

In this paper, we assumed the particular molecular structure of the pore states similar to that proposed earlier in a number of papers [16,31,32], also reviewed in [35]. We utilized the continuum theory of elasticity of lipid membranes to analyze both local and non-local mechanisms of pore formation by AMPs, corresponding to low and high concentration regimes, respectively. In earlier works, the energy of discreet (meta)stable states rather than a continuous trajectory of the pore formation process was calculated [16,36,37]. The difference of the energies of the (meta)stable states provides only the lower estimate of the energy barriers of pore formation process. Using the geometric parameters of the peptides and known values of the elastic parameters of lipid membranes, we calculated continuous trajectories of pore formation starting from the intact planar bilayer through the hydrophobic defect to the transversal pore. The continuous trajectories allowed us to obtain an upper estimate of the energy barriers of pore formation. In the local mechanism, a small pore, permeable to protons but not to large ions or calcein, can be formed by as few as two AMP molecules partially inserted into an almost tensionless membrane. Below, we refer to such a pore as proton pore or $H^+$-pore. Within the non-local mechanism, an intensive formation of large and relatively long-living pores is predicted to occur at a sufficiently high (above threshold) concentration of AMPs. Very conditionally, the local mechanism may be related to the graded mechanism of LUV permeabilization, while non-local mechanism has some similarity points with the all-or-none mechanism [31]. The results of the theoretical modeling are related to the experimental observations described in the companion paper [38].

## II. THEORY

### A. Elastic energy functional

We calculated the energy of elastic deformations of the membrane in the framework of the theory of elasticity of lipid membranes originally developed by Hamm and Kozlov [39,40]. Further, the theory was generalized to include additional deformation modes [41–43]. In the Hamm and Kozlov theory, an average orientation of lipid molecules is characterized by a vector field of unit vectors $\mathbf{n}$, called directors. This vector field is defined on the surface lying inside the lipid monolayer in the region of junction of polar lipid heads and hydrophobic chains; the surface is referred to as the neutral surface [44]. The shape of the neutral surface is characterized by the vector field of its unit normal, $\mathbf{N}$. All deformations are deemed small and the energy is a quadratic function of them. We account for the following deformations: 1) splay, characterized by the divergence of the director along the neutral surface, div($\mathbf{n}$); 2) tilt, characterized by the tilt-vector $\mathbf{t} = \mathbf{n} - \mathbf{N}$; 3) Gaussian splay, which in some tangential Cartesian coordinate system $Oxy$ can be written as $K = \dfrac{\partial n_x}{\partial x}\dfrac{\partial n_y}{\partial y} - \dfrac{\partial n_x}{\partial y}\dfrac{\partial n_y}{\partial x}$, where $n_x$, $n_y$ are the corresponding projections of the director; 4) lateral stretching, characterized by the relative change in the area per lipid molecule at the neutral surface, $\alpha = (a - a_0)/a_0$, where $a$, $a_0$ are current and initial areas per molecule, respectively; 5) lateral tension $\sigma$, which accounts for the work of external forces against an increase in the area of the neutral surface in the course of deformations; 6) twist, characterized by $\mathbf{rot}(\mathbf{n})$. The surface density of the elastic deformations energy of a lipid monolayer is written as [22]:

$$w = \frac{k_c}{2}\Big[\operatorname{div}(\mathbf{n}) + J_0\Big]^2 - \frac{k_c}{2}J_0^2 + \frac{k_t}{2}\mathbf{t}^2 + \frac{k_a}{2}\big(\alpha - \alpha_0\big)^2 + \\ + \frac{\sigma}{2}\Big[\mathbf{grad}(H)\Big]^2 + k_G K + \frac{k_{rot}}{2}\Big[\mathbf{rot}(\mathbf{n})\Big]^2,$$
(1)

where $k_c$, $k_t$, $k_a$, $k_G$, $k_{rot}$ are elastic moduli of splay, tilt, lateral stretching, Gaussian splay, and twist, respectively; $J_0$ is the spontaneous curvature of the lipid monolayer; $\alpha_0 = \sigma/k_a$ is the spontaneous stretching caused by the constant lateral tension $\sigma$; in the Cartesian coordinate system, where the $Oz$ axis is perpendicular to the membrane plane, $H = H(x, y)$ is the function determining the $z$-coordinates of points of the neutral surface. The bulk modulus of lipid membranes is very high, $\sim 10^{10}$ J/m$^3$ [45]. For this reason, we considered the hydrophobic part of lipid monolayers volumetrically incompressible. The functional of the elastic energy of the lipid bilayer is given by the sum of the functionals of its constituent monolayers [46]. See Supplementary Materials for details of calculations.

According to Marcelja's theory [47], the energy of the cylindrical hydrophobic cavity of radius $R_0$ and length $H_h$ coaxial with the $Oz$-axis and filled with water can be obtained in the following form:

$$W_{hyd} = \sigma_0 \left(2\pi R_0 H_h\right) \frac{I_1\left(\dfrac{R_0}{\xi_h}\right)}{I_0\left(\dfrac{R_0}{\xi_h}\right)}, \tag{2}$$

where $I_0$, $I_1$ are modified Bessel functions; $\xi_h \approx 1$ nm is the characteristic length of hydrophobic interactions; $\sigma_0 \approx 40$ mN/m is the surface tension of the planar macroscopic water-lipid tails interface [39,48]; $(2\pi R_0 H_h)$ is the area of the side-surface of the cylinder. This energy is added to the energy of membrane deformations Eq. (S4).

Eq. (2) is obtained based on Marcelja's formalism [47], which assumes different properties of water in the vicinity of hydrophobic walls as compared to the bulk water. The difference is characterized by the order parameter $\eta$, which is assumed zero in the bulk. Minimization of the energy functional written in the Ginzburg-Landau form yields the Euler-Lagrange equation for the order parameter: $\xi_h^2 \Delta \eta = \eta$. In the case of cylindrical symmetry, i.e., if the hydrophobic cavity is assumed to be an infinitely long cylinder without any edge effects, the Euler-Lagrange equation with the boundary conditions $|\eta| < \infty$, $\eta(R_0) = \eta_0$ can be solved analytically. If we substitute the solution into the energy functional and subsequently integrate it, we obtain the expression given by Eq. (2). An obvious shortcoming of this algorithm is that we assume that the hydrophobic cylinder is of infinite length. The energetic cost of the hydrophobic cavity was calculated much more accurately in [49]. In this paper, the energy of a non-cylindrical (although rotationally symmetric) hydrophobic cavity was obtained numerically with explicit accounting for the finite size of the cavity and presence of polar lipid heads near the cavity. However, the calculation required imposing the boundary condition on the order parameter $\eta$ at the polar surface of lipid heads; the authors have put the condition $\eta = 0$. Physically, the order parameter $\eta$ characterizes the difference of some physical property of water (e.g., average number of H-bonds per unit volume, average density, etc.) near a hydrophobic surface and in the bulk. The boundary condition $\eta = 0$ at the surface of polar lipid heads means that the properties of water near polar lipid heads are the same as in the bulk (where $\eta$ is also zero, by definition). This assumption does not seem entirely valid, as polar lipid heads orient the water thus altering its physical properties, e.g. its dielectric constant. A more realistic boundary condition should be some non-zero value of the order parameter at the surface of polar lipid heads. However, this value is not known. Thus, a formally more accurate calculation of hydrophobic cavity energy, proposed in [49], should rely on unknown boundary value of the order parameter. For this reason, we used expression Eq. (2) for calculating the hydrophobic cavity energy. We realize that it is not quite accurate, but it is not obvious that more detailed consideration (e.g., similar to the one presented in [49]) improves the accuracy.

The elastic energy functional should be supplemented by appropriate boundary conditions. In general, the monolayer director and the neutral surface are assumed to be continuous everywhere except for the region of the hydrophobic cylinder in the hydrophobic defect state. The deformations should be finite (and small) and should decay far from the pore and amphipathic peptides. In the non-local mechanism of pore formation, we impose only these general conditions along with the specific boundary conditions (S6), (S7) on the hydrophobic defect. In the local mechanism of H$^+$-pore formation, amphipathic peptides are explicitly considered as solid membrane-deforming inclusions.

To obtain the spatial distribution of deformations and elastic energy, one should minimize the functional of the total energy in the form (S4) written for the lipid bilayer subjected to the boundary conditions formulated above. However, the minimization cannot be performed analytically as the Euler-Lagrange equations derived in the course of the functional variation are partial differential equations that cannot be solved under the imposed boundary conditions. Thus, functional (S4) was minimized numerically using the finite element method, as in [22,50,51]. Briefly, the neutral surface of each monolayer was divided into elementary triangles. The mesh was not homogeneous: the size of the triangles decreased in the vicinity of each peptide and pore. Inside each triangle the deformations were substituted by their linear interpolants, and the surface density of the elastic energy was analytically integrated over the area of the triangle. Such elementary energies depend quadratically on the deformations values in the vertices of elementary triangles. The elementary energies were summed yielding the total elastic energy of the membrane depending on the values of deformations in the vertices of elementary triangles. The total energy was minimized with respect to these values, except for the vertices located at the peptide or pore boundary, as these values are determined by the boundary conditions. The total size of the membrane was adjusted in a way ensuring that the boundary of the membrane was at least 25 nm away from the peptides or pore, so it exceeded the characteristic length of deformation decay (~6 nm) [22,50].

The membrane with a circular hydrophobic defect, pore, or ring-like configuration of the peptides possesses a rotational symmetry. This symmetry allows us to minimize analytically the elastic energy functional (S4). Variation of the functional with respect to the functions characterizing the state of lipid bilayer yields five Euler-Lagrange equations (S10); they were solved analytically (see Supplementary Materials).

Energy functional (S9) for lipid bilayer was derived under the assumption of small deformations. Formally, the director should deviate but slightly from the unit normal vector to the surface of the flat undeformed bilayer. However, in the equatorial plane of the pore, the director is almost perpendicular to the unit normal vector. Thus, the energy functional (S9) is inapplicable in the vicinity of the pore equator. However, the deformations at the pore edge can still be considered small if a vertical cylindrical monolayer coaxial with the axis perpendicular to the membrane plane is considered as the reference state instead of the planar horizontal bilayer. This consideration allows to formally apply the linear theory of elasticity. The zone of the vertical cylindrical monolayer is conjugated with the bilayer region of the membrane based on the continuity of neutral surfaces and directors. The spatial distribution of deformations in the vertical monolayer region was obtained analytically (see Supplementary Materials). The vertical monolayer is conjugated with the upper and lower monolayers of the bilayer part of the membrane along two circles [Fig. 1(c)]. The total energy was numerically minimized with respect to the coordinates of conjugation of the horizontal bilayer and vertical monolayer parts of the membrane, as well as with respect to the coordinates of the ends of the hydrophobic cylinder using the gradient descent method [25]. The symbolic computations and optimization programs are available upon request.

## III. RESULTS

### A. Parameters of the system

To obtain quantitative results, we used the following values of the elastic parameters, typical for dioleoylphosphatidylcholine (DOPC) membranes (per monolayer): splay modulus $k_c$ = 10 $k_B T$ ($k_B T \approx 4 \times 10^{-21}$ J) [52]; tilt modulus $k_t$ = 40 mN/m [39]; lateral stretching modulus $k_a$ = 133 mN/m [52]; Gaussian splay modulus $k_G = -0.3k_c$ [53]; spontaneous curvature $J_0 = -0.091$ nm$^{-1}$ [54,55]; thickness of the hydrophobic part of the monolayer $h$ = 1.45 nm [25,52], twist modulus $k_{rot} = k_c/2$ [56]. The peptide diameter $2r_0$ = 1.3 nm; the peptide length $L_p$ = 5 nm [see [Fig. S1(a)]. When two peptide molecules form a ring-like structure, the radius of the peptide ring midline is assumed $R_p = 2L_p/(2\pi) \approx 1.6$ nm. The actual length of magainin is about 3.5 nm (23 amino acid residues, 0.15 nm per residue in $\alpha$-helix) — shorter than the supposed length $L_p$ = 5 nm. In the present work, we mainly focused on elastic deformations induced by amphipathic peptides. As deformations do arise in the vicinity of two ends of the helix, we used the so-called effective "deformational" length of helical peptides, which was assumed to be somewhat longer than their actual length [22].

### B. Pore formation in a symmetric lipid membrane without peptides

First, we considered the process of lipidic pore formation in a symmetric membrane with identical lateral tensions in two monolayers. The continuous trajectory of the pore formation process was described by two coordinates: the pore lumen radius, $R_0$, and the length of the hydrophobic cylinder, $L$. Due to symmetry, the monolayer interface remained flat for all values of $R_0$ and $L$ (Fig. S2). The energy landscapes $W(R_0, L)$ are presented in Fig. S3 for different values of the lateral tension per monolayer (from 0.01 to 4.5 mN/m). At each fixed value of the lumen radius, the pore energy has a minimum at some length of the hydrophobic cylinder (Fig. S3). The minima determined the optimal trajectory $L_0(R_0)$ of pore formation. By substituting the optimal trajectory $L = L_0(R_0)$ into the pore energy $W(R_0, L)$ we reduced the independent coordinates to the pore lumen radius, $R_0$ only, as $W_0(R_0) = W(R_0, L_0(R_0))$. Thus, we obtained optimal trajectories connecting the states of intact bilayer ($L = 2h$, $R_0 = 0$) and pore ($L = 0$, $R_0 > 0$) along the energy surface $W(R_0, L)$. The dependences of pore formation energy along the optimal trajectories $L = L_0(R_0)$ are presented in Fig. 3(a) for different values of the symmetric lateral tension of the membrane monolayers.

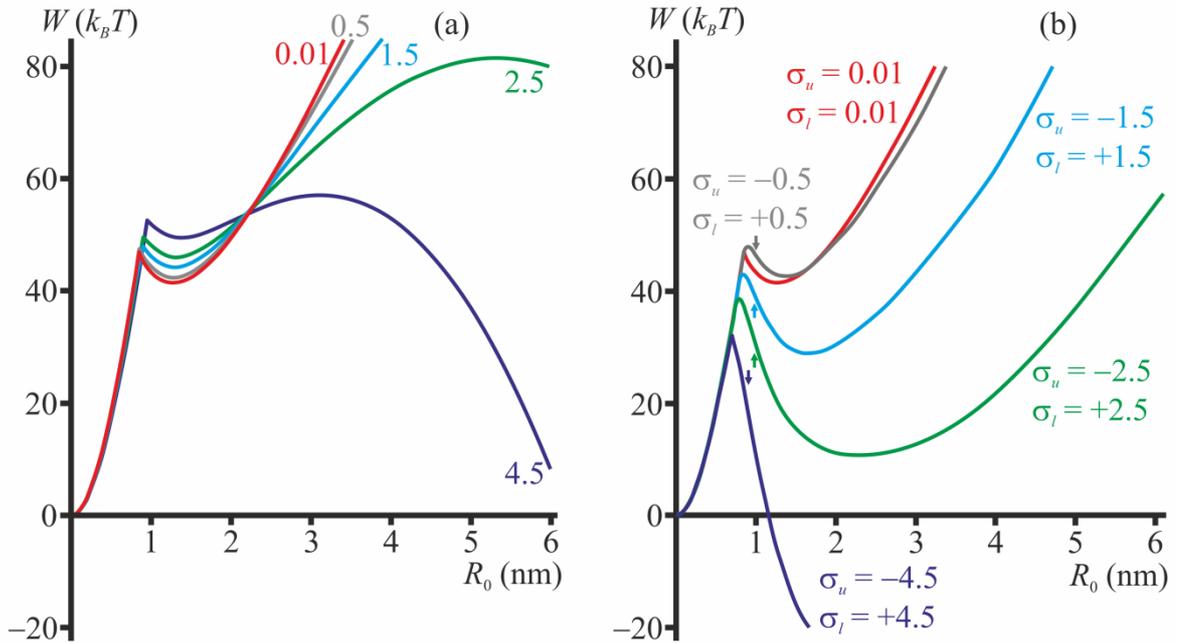

FIG. 3. Optimal trajectories of the lipidic pore formation $W(R_0)$ in the membrane whose monolayers are subjected to: (a) — equal lateral tensions; (b) — asymmetrically applied lateral pressure/tension. The pressure/tension (in mN/m) is indicated at each curve. The vertical arrows in panel (b) indicate the radius at which the length of the hydrophobic cylinder becomes equal to zero and a hydrophilic pore is formed. The red curves corresponding to the symmetric lateral tension $\sigma_u = \sigma_l = 0.01$ mN/m coincide in panels (a) and (b).

Generally, at any positive lateral tension there are two energy barriers, i.e., there are two maxima of the dependence $W_0(R_0)$, separated by a local minimum. The energy barrier at the lumen radius $R_0 \approx 1$ nm corresponds to the transformation of the hydrophobic defect [Fig. 1(c)] into the hydrophilic pore [Fig. 1(b)]. This occurs when the energy of the hydrophobic defect ($L \neq 0$) becomes equal to the energy of the hydrophilic pore ($L = 0$) of the same radius. The corresponding transition is indicated by horizontal double-directed arrows in Fig. S2; the transition is accompanied by a sharp drop in the length of the hydrophobic cylinder from $L \approx 2$ nm to $L = 0$. The height of this energy barrier slightly increases when the lateral tension increases [Fig. 3(a)]; its position, however, is weakly dependent on the lateral tension. At the radius $R_0 \approx 1.5$ nm the pore energy has a local minimum, meaning that the hydrophilic pore can be metastable and can have a finite lifetime. The second energy barrier is located at the pore lumen radius $R^* \approx \gamma/\sigma$, which is strongly dependent on the lateral tension; this radius decreases fast with increasing tension [Fig. 3(a)]. Overcoming the second energy barrier leads to an irreversible rupture of the membrane, since the pore energy monotonically decreases with increasing pore radius for $R_0 > R^*$, as qualitatively described by the Derjaguin-Gutop model. At the lateral tension $\sigma_u = \sigma_l \approx 4.5$ mN/m the heights of the two energy barriers are almost tie and equal to about 52.5 $k_B T$ [Fig. 3(a)]. The minimal energy barrier of the metastable hydrophilic pore formation is about 46.5 $k_B T$ (e.g., at a low lateral tension of $\sigma_u = \sigma_l = 0.01$ mN/m).

## C. Pore formation in peptide-free membrane with asymmetric lateral pressure/tension

Further, we considered a membrane with lateral tension in the lower monolayer ($\sigma_l > 0$) and lateral pressure in the upper monolayer ($\sigma_u < 0$). The tension and pressure were assumed to be generated by one-sided peptide adsorption in high concentration. However, peptide molecules were not considered explicitly, their action was limited to the generation of the lateral pressure/tension.

For the membrane to be stable, the total tension ($\sigma_u + \sigma_l$) is required to be positive. To ensure that this condition holds, we considered the cases of $\sigma_u = -\sigma_0 + \sigma_e$, $\sigma_l = \sigma_0 + \sigma_e$, where $\sigma_0 = 0.5, 1.5, 2.5, 4.5$ mN/m, and $\sigma_e = 0.01$ mN/m was a low additional tension having almost no effect on the pore energy and deformation distribution, but ensuring that ($\sigma_u + \sigma_l$) $= 2\sigma_e = 0.02$ mN/m $> 0$. In the case of such asymmetrically applied lateral pressure/tension, the membrane shape became asymmetric as well: monolayer interface deviated from a plane (Fig. S4). It is energetically favorable to maximize the surface of the upper (compressed) monolayer and simultaneously minimize the surface of the lower (stretched) monolayer (Fig. S4). At the stage of the small radius hydrophobic defect, the optimal surface is achieved when the monolayer interface bends downwards [Fig. S4(a)], while for the defect or pore of a larger radius, the monolayer interface bends upwards [Fig. S4(b), (c)]. In some range of hydrophobic defect radii both types of the membrane shape can occur as the total energy has a local minimum in both configurations. Thus, the shape of the membrane with asymmetric lateral pressure/tension may exhibit bistability. We explored both branches of the membrane shapes, and considered the shape of minimal total energy at each radius. Note that the energies of two coexisting shape configurations (bent upwards or downwards) differed, but only slightly, typically by less than 3 $k_BT$.

In the plots of the membrane shape in the vicinity of the pore edge, there are discontinuities in the gradient of the lipid monolayers' neutral surface (Fig. S4). Formally, this is not a problem as the neutral surface and the director remain continuous. The surface gradient discontinuity arises at the junctions of "horizontal" and "vertical" monolayers. Of course, this is the same lipid monolayer, and there is no physical reason for the discontinuity; the reason is purely mathematical. As we use the linear theory of elasticity, we have to keep deformations small. This is partially achieved by assuming different reference states for lipid monolayer parts located close to the equatorial plane of the pore ("vertical" monolayer) and at some distance from the pore equator ("horizontal" monolayer). The reference state for the "horizontal" monolayer is a plane, while the reference state for the "vertical" monolayer is a cylinder. The deviations of the respective monolayers from the corresponding reference states are assumed to be small, which allows to formally apply the linear theory of elasticity. However, at the points where the "horizontal" and "vertical" monolayers meet, the deformations with respect to both reference states are not small. The deformation energy of the monolayer region adjacent to the joint points is calculated inaccurately, and this leads to the appearance of the angle between neutral surfaces of the "vertical" and "horizontal" parts of the lipid monolayer. The energy penalty for the appearance of the angle is imposed by the tilt deformation: tilt appears unavoidably in the junction region of "vertical" and "horizontal" monolayers, as the director is continuous at the point of junction, while the unit normal (related to the gradient of the surface) is discontinuous. The angles (as well as the calculated elastic energy of the pore edge) can be decreased by introducing the third region of the lipid monolayer between the "vertical" and "horizontal" ones, with its own reference state. However, the introduction of the third region would complicate the calculations substantially. If we consider only two regions ("vertical" and "horizontal") of a lipid monolayer, we overestimate the elastic energy of the pore. Imposing the continuity condition on the surface gradients will lead to an even greater overestimation of the energy.

The energy landscape $W(R_0, L)$ of the hydrophobic defect and the hydrophilic pore in the membrane with asymmetrically applied lateral pressure/tension is shown in Fig. S5 for different values of the pressure/tension. The energy has a global minimum at each fixed value of the pore lumen radius $R_0$ (Fig. S5). The minima determine the optimal trajectory $L_0(R_0)$ of pore formation. The substitution of the optimal trajectory $L = L_0(R_0)$ into the pore energy $W(R_0, L)$ allows us to obtain the optimal trajectories $W_0(R_0) = W(R_0, L_0(R_0))$ connecting the states of intact bilayer ($L = 2h$, $R_0 = 0$) and pore ($L = 0$, $R_0 > 0$) along the energy surface $W(R_0, L)$. Dependences of the pore formation energy along the optimal trajectories $L = L_0(R_0)$ are presented in Fig. 3(b) for different values of lateral pressure/tension applied asymmetrically to membrane monolayers. In these cases, the single energy barrier at $R_0 \approx 1$ nm remains in a reasonable range of pore radii. The barrier corresponds to the transformation of the hydrophobic defect into the hydrophilic pore. At the top of the barrier, some additional energy of the order of several $k_BT$s should be spent to proceed from the hydrophobic defect of a long length (e.g., $L \approx 2$ nm) to the hydrophobic defect of a short length (e.g., $L \approx 0.7$ nm), see gray curve in Fig. S5(a), orange curve in Fig. S5(b), green curve in Fig. S5(c). Both types of defects are characterized by local minima of the pore energy, but these minima are separated by a local maximum (Fig. S5). For the dependences $W(R_0)$ shown in Fig. 3(b), the formal heights of the energy barriers at $R_0 \approx 1$ nm are 47.8 $k_BT$ ($\sigma_l = -\sigma_u = 0.5$ mN/m), 42.9 $k_BT$ ($\sigma_l = -\sigma_u = 1.5$ mN/m), 38.6 $k_BT$ ($\sigma_l = -\sigma_u = 2.5$ mN/m), and 32.2 $k_BT$ ($\sigma_l = -\sigma_u = 4.5$ mN/m). However, if the local maxima in the energy landscapes $W(R_0, L)$ are taken into account, the total heights of the energy barriers would be 50.8 $k_BT$, 46.4 $k_BT$, 42.5 $k_BT$, and 35.4 $k_BT$, respectively. Such energy dependences (two minima of equal energy separated by a maximum) reflect switching from the downward bent to the upward bent membrane shapes corresponding to long and short lengths of the hydrophobic defect, respectively (see Fig. S4); this switch requires surmounting the local energy barrier (Fig. S5).

Overcoming the energy barrier leads to the formation of a metastable pore characterized by a local minimum of the pore energy $W(R_0)$ at $R_0 \geq 1.5$ nm [Fig. 3(b)]. In the case of symmetric lateral tension, the depth of the local minima decreased with increasing tension [Fig. 3(a)], while the positions of the minima remained almost unaltered, $R_0 \approx h \approx 1.4$ nm. In the asymmetric case of applied lateral pressure/tension, the corresponding minima became deeper at increasing pressure/tension [Fig. 3(b)] and they shifted towards larger radii. This means that in this case the metastable pores should be larger and much more long-living, as their closure requires overcoming a higher energy barrier. For comparison, the energy barriers of metastable pore closure ($R_0 \rightarrow 0$) are 5.6 $k_BT$, 5.1 $k_BT$, 14.0 $k_BT$ and 27.7 $k_BT$ for $\sigma_l = \sigma_u = 0.01$ mN/m, $\sigma_l = -\sigma_u = 0.5$ mN/m, $\sigma_l = -\sigma_u = 1.5$ mN/m and $\sigma_l = -\sigma_u = 2.5$ mN/m, respectively. At a sufficiently high lateral pressure/tension [e.g., $\sigma_l = -\sigma_u = 4.5$ mN/m, blue curve in Fig. 3(b)], the pore energy has global rather than a local minimum at a finite non-zero radius $R_0$, implying that the ground state in this case corresponds to a perforated membrane, while the defect-free membrane ($R_0 = 0$) becomes metastable.

## D. Ground state of a symmetric membrane with inserted amphipathic peptides

We considered two molecules of amphipathic peptides [Fig. S1(a)] partially inserted [Fig. S1(b)] into the upper monolayer of the horizontal lipid bilayer. Recently, we have shown that the ground state of such system is the so-called registered configuration of peptides as shown in Fig. 4(a) [22]. Figure 4(a) illustrates the distribution of the membrane elastic stress in the optimal registered configuration ($L \approx 6$ nm) for peptides located in the same lipid monolayer. In the middle between the peptides, the membrane is under the stress of appreciable amplitude. We thus hypothesized that the formation of the pore is most probable in this region of the membrane. The dependence of the membrane elastic energy on the distance between axes of two parallel registered amphipathic peptides located in the same lipid monolayer or in the opposing monolayers is shown in Fig. 4(b). The plots are built for the projection of the boundary director at the peptide boundary $|n_0| = 0.4$ [see boundary conditions, Eq. (S8)]; the elastic energy scales as $n_0^2$.

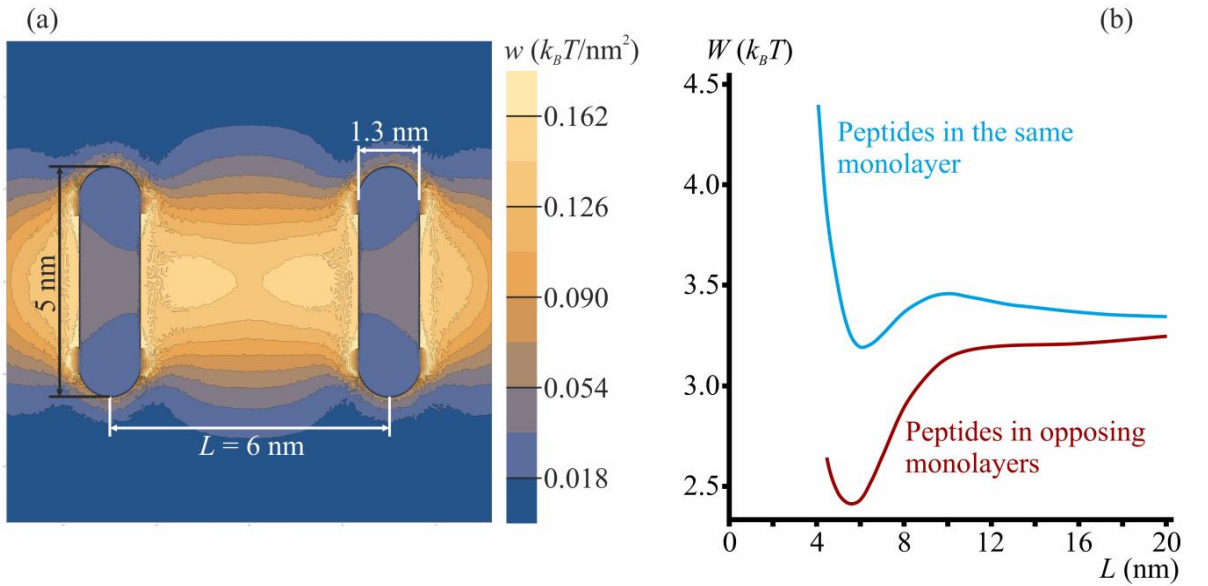

FIG. 4. The energy of membrane deformations in the registered configuration of two amphipathic peptide molecules. (a) — The registered configuration of two amphipathic peptides. The optimal distance between peptide axes is about 6 nm. The spatial distribution of the density of the membrane elastic energy is shown for the case of peptides located in the same (upper) monolayer. (b) — The dependence of the membrane elastic energy on the in-plane distance between the axes of two parallel registered amphipathic peptides located in the same lipid monolayer (blue curve) or in opposing monolayers (red curve). The projection of the boundary director $|n_0| = 0.4$.

Both dependences have the global minimum at the distance $L \approx 6$ nm between the axes of the peptides. However, the minimum is substantially deeper when the peptides are located in opposing monolayers [Fig. 4(b)].

**E. Hydrophobic defect formed between two amphipathic peptides**

The formation of the hydrophilic pore in a membrane is generally thought to occur via an intermediate state of the so-called hydrophobic defect [26,27]. The hydrophobic defect is the transmembrane cylinder whose side-walls are lined with hydrophobic chains of lipid molecules. If the length $H_h$ of the cylinder is less than the thickness of the hydrophobic zone of the bilayer, $2h$, then elastic deformations should arise in the vicinity of the hydrophobic cylinder boundary. Thus, the total energy of the hydrophobic defect includes the hydrophobic interactions energy, Eq. (2), and the elastic deformations energy. The total energy is determined by the radius $R_0$ and the length $H_h$ of the hydrophobic cylinder.

The optimal registered configuration of peptides [Fig. 4(a)] lacks a rotational symmetry. Thus, in general, the shape of the hydrophobic defect formed in the middle between the peptides is rather elliptic-like than circular. However, for small enough radii of the defect, the energy difference between circular and elliptic hydrophobic defects should be negligible. We thus considered the formation of a circular (cylindrical) hydrophobic defect in the middle between two peptides standing in the optimal registered configuration. In addition, we took into account that the typical persistence length of an α-helix is of the order of nanometers [57], i.e., it is shorter than the assumed peptide length $L_p$. This means that helical peptides are rather flexible and can bend substantially at the expense of relatively low thermal energy (~1 $k_B T$). We allowed the peptides to bend around the cylindrical hydrophobic defect to form a ring-like structure (Fig. 5, inset), assuming effectively zero bending modulus of an $\alpha$-helix. The total energy of cylindrical hydrophobic defects of different radii is illustrated in Fig. 5 for two bent and two straight peptides.

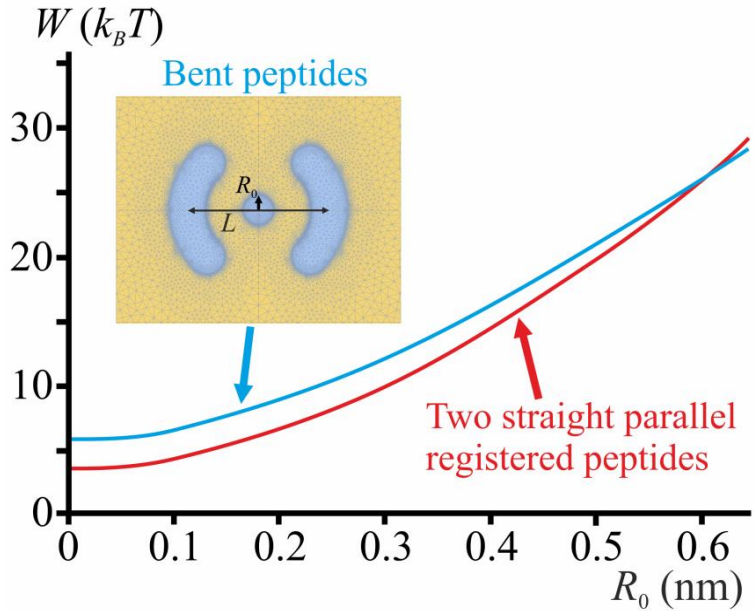

FIG. 5. Energy dependences of the cylindrical hydrophobic defect on its radius. The defect is formed in the middle between two amphipathic peptides that are allowed to bend around the defect. After overcoming some radius ($R_0 \geq 0.6$ nm), the configuration of the peptides bent into a ring-like structure (blue curve) becomes more energetically favorable than the initial optimal registered configuration [Fig. 4(a)] of straight parallel peptides (red curve). Bent peptides with a circular hydrophobic defect formed in between them are shown in the inset. The inhomogeneous triangular lattice used for the numerical minimization of the elastic energy is shown on the yellow background that indicates the surface of the lipid monolayer; the blue color corresponds to the regions where the lipid monolayer neutral surface does not exist (two bent peptides and a circular hydrophobic defect in between them). In the inset, $L = 6$ nm.

At small radii ($R_0 < 0.6$ nm), the optimal configuration of the system coincides with the registered configuration of the peptides. However, upon overcoming some radius ($R_0 \geq 0.6$ nm), the configuration of the peptides bent into a ring-like structure becomes more energetically favorable than the initial optimal registered configuration of straight linear peptides (Fig. 5). This can be interpreted as follows. The formed hydrophobic defect causes the amphipathic peptides to bend into a ring-like structure with a rotational symmetry. The intensive re-orientation of lipids at the pore edge and the associated top of the energy barrier of the hydrophilic pore formation corresponds to the radius of the hydrophobic defect of about $R_0 \geq 0.8$ nm [25]. In this range of radii, bent peptides are more energetically favorable than the straight registered ones (Fig. 5). Thus, further we considered ring-like configurations of peptides possessing a rotational symmetry in the whole range of radii of the hydrophobic defect, as this allows simplifying the calculations. According to calculation results, in the most interesting range of radii (e.g., $R_0 \geq 0.6$ nm) the peptides do form ring-like structures).

**F. Pore formation by amphipathic peptides in rotationally symmetric configurations**

We analyzed the energy landscape of pore formation induced by explicitly considered amphipathic peptides under the assumption of the rotational symmetry of the system. As a reaction coordinate, we used the tilt angle of the peptide ring, $n_t$ [Fig. 6(b)]. Due to symmetry, when peptides are located far from each other, the average director of lipids adjacent to the long sides [left and right sides in Fig. S1(a)] of the peptide outline coincides with the unit normal vector of the plane of the undisturbed membrane[Fig. S1(b)]. The same condition holds at the stage of the hydrophobic defect in the rotationally symmetric configuration where $n_t = 0$ [Fig. 6(a)]. When peptides occupy the equatorial plane of the hydrophilic pore, the average director of the adjacent lipids is horizontal, i.e., it is perpendicular to the unit normal vector of the membrane plane, and $n_t = 1$ [Fig. 6(c)]. Thus, the process of pore formation can be parameterized by the coordinate $n_t$ from the initial optimal registered configuration and the hydrophobic defect of a small radius [Fig. 6(a)] to the lipid-peptide pore[Fig. 6(c)]. We assumed that the peptides did not shift laterally, i.e., at any value of the tilt angle $n_t$, they formed the ring-like structure of the constant radius at the peptide midline, $R_p = 2L_p/(2\pi) \approx 1.6$ nm (Fig. 6).

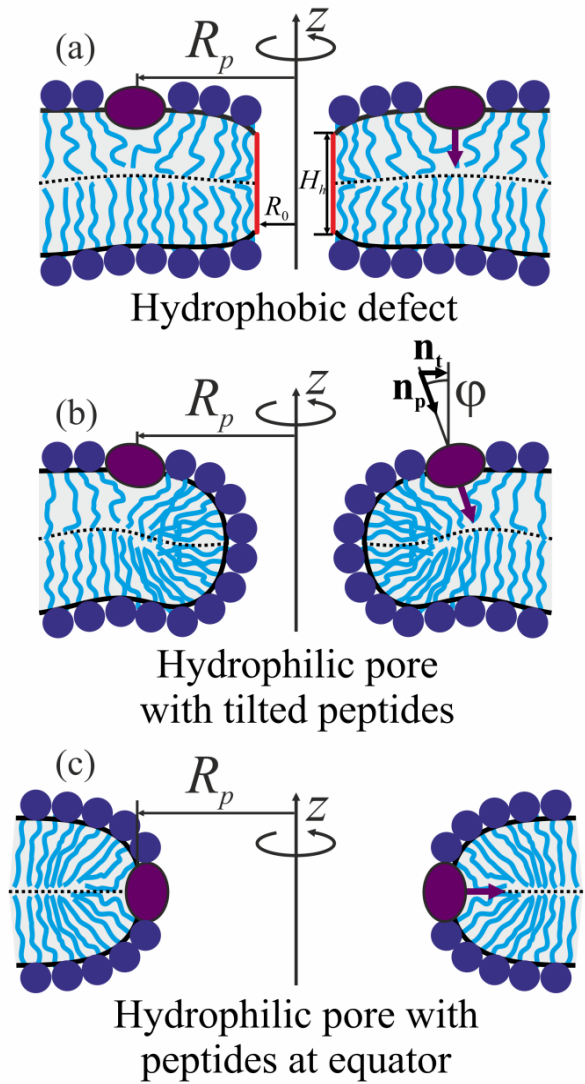

(a)

Hydrophobic defect

(b)

Hydrophilic pore
with tilted peptides

(c)

Hydrophilic pore with
peptides at equator

FIG. 6. The system configurations for different values of the tilt angle $n_t$ of the peptide ring. The value $n_t \approx 0$ corresponds to the initial optimal registered configuration (a); the value $n_t = 1$ corresponds to the toroidal lipid-peptide pore with the peptides lining the pore equator (c); intermediate values $0 < n_t < 1$ correspond to intermediate configurations of the hydrophobic defect of a large radius or hydrophilic pore (b).

We sequentially fixed different values of $n_t$, varied the radius of the hydrophobic defect $R_0$ and optimized the system energy with respect to the length $H_h$ of the hydrophobic defect, thus, obtaining the energy surface $W(n_t, R_0)$ illustrated in Fig. S6. We considered different values of the boundary director projection at the peptide boundaries, $n_0$ [see Eq. (S8)] [$|n_0| = 0.3$, Fig. S6(a), (d); $|n_0| = 0.4$, Fig. S6(b), (e); $|n_0| = 0.5$, Fig. S6(c), (f)], and different values of the membrane lateral tension [low tension $\sigma = 0.1$ mN/m (per monolayer), Fig. S6(a)–(c), and high tension $\sigma = 4$ mN/m (per monolayer), Fig. S6(d)–(f)]. We did not calculate the dependence $W_0(n_t)$ for the case of exactly zero lateral tension, although the theory is applicable in this case. However, the analytical solutions of the Euler-Lagrange equations (S10) in the cases of exactly zero lateral tension and low (but positive) lateral tension are based on different mathematical functions. E.g., in the case of exactly zero lateral tension the shape of monolayer interface is described by logarithms and polynomials of the third order, while in the case of any positive lateral tension the shape is described by the Bessel functions. Thus, we did not explicitly consider the case of exactly zero lateral tension, as this consideration requires independent derivation of all equations; we used a low non-zero tension $\sigma = 0.1$ mN/m instead. Quantitatively, the case $\sigma = 0.1$ mN/m differs but slightly from the case of exactly zero tension: such a low lateral tension (0.1 mN/m) has almost no effect on the elastic energy and characteristic lengths of the deformations. Typical values of the lateral tensions measured in cell membranes are in a range from 0.01 to 0.3 mN/m [58].

The projection of the boundary director induced by insertion of an amphipathic helix into a lipid monolayer by the endocytic protein dynamin-1 was estimated to be about $|n_0| = 0.4$ [59]. Amphipathic peptides inserted into a lipid membrane cause thinning of the lipid bilayer, which was determined by using X-ray scattering [12,30]. Recently, the thinning effect has been described in the framework of the developed elastic model [60]. We fitted the theoretical results to the experimental data obtained for magainin [12] and melittin [30], with $n_0$ being the only fitting parameter. The predicted values of the boundary directors are: $|n_0| = 0.42$ for melittin and $|n_0| = 0.48$ for magainin [60]. The minima of the dependencies $W(R_0)$, shown as circles in Fig. S6 determine the optimal trajectory $R_0(n_t)$ of pore formation by two molecules of amphipathic peptides. The energy surface $W(n_t, R_0)$ has a saddle point, which is the highest minimum of the dependencies $W(R_0)$ at different fixed values of $n_t$. The saddle points on the plots in Fig. S6 correspond to the top of the energy barrier of the pore formation process.

The pore energy dependence $W(n_t) = W(n_t, R_0(n_t))$ along the optimal trajectory $R_0(n_t)$ of pore formation is presented in Fig. 7 for low lateral tension [Fig. 7(a)] and high lateral tension [Fig. 7(b)].

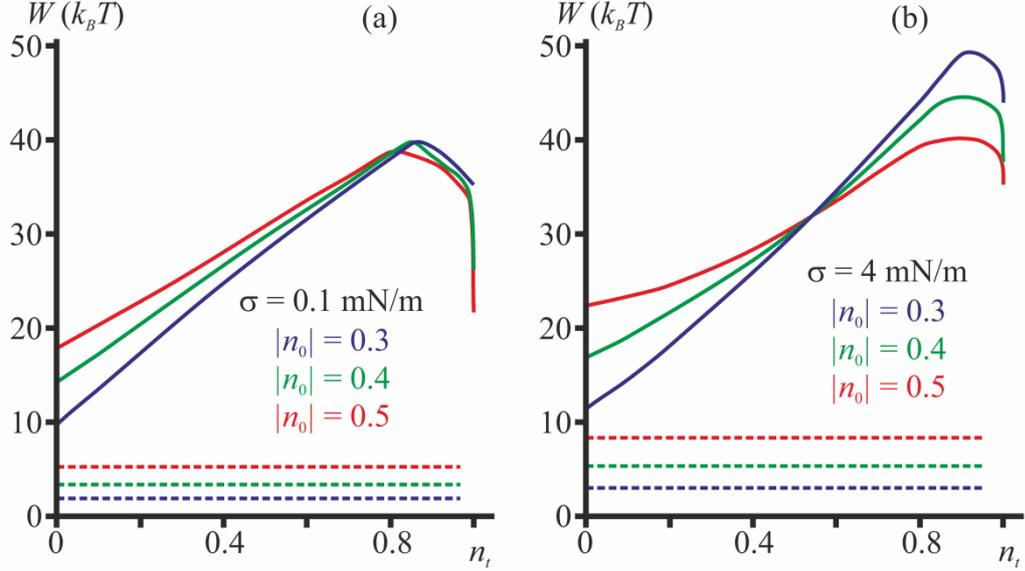

FIG. 7. The pore energy dependences $W(n_t) = W(n_t, R_0(n_t))$ along the optimal trajectories $R_0(n_t)$ of pore formation in the case of low lateral tension [$\sigma = 0.1$ mN/m per monolayer, panel (a)] and high lateral tension [$\sigma = 4$ mN/m per monolayer, panel (b)]. The dependences are built for different values of the boundary director projection ($|n_0| = 0.3$ blue curves; $|n_0| = 0.4$ green curves; $|n_0| = 0.5$ red curves). Horizontal dashed lines correspond to the energy of elastic deformations induced by two far-separated peptide molecules.

Each dependence $W(n_t)$ has a global maximum at $n_t = 0.8$–$0.9$ (Fig. 7). The difference between the energy at the maximum and the energy of the elastic deformations induced by two far separated peptide molecules (dashed horizontal lines in Fig. 7) determines the height of the energy barrier of the pore formation process. Generally, the energy barrier decreases as the absolute value of the projection of the boundary director $|n_0|$ increases. The boundary director value determines the curvature induced by the inserted peptide [compare Fig. S1(d) and Fig. S1(e)]. A larger $n_0$ corresponds to a higher positive curvature of the membrane monolayer into which the peptide is inserted. The pore edge is also characterized by a positive average curvature. Thus, the energy of the pore edge with the peptides incorporated into its equator is lower for larger $n_0$, i.e., for higher positive curvature induced by the peptide (compare the energy values for different $n_0$ at $n_t = 1$ in Fig. 7). Besides, the energy barrier of pore formation is determined as the difference between the pore energy at its maximum ($n_t \approx 0.9$) and the energy of elastic deformations induced by two far-separated peptide molecules in a planar membrane (dashed horizontal lines in Fig. 7). The latter energy increases with increasing $n_0$. This also contributes to the decrease in the energy barrier when $n_0$ increases. The influence of the lateral tension is less obvious: the energy barrier increases with an increasing lateral tension for a small projection of the boundary director ($|n_0| = 0.3$ and $|n_0| = 0.4$) and the energy barrier decreases with an increasing lateral tension for a large projection of the boundary director ($|n_0| = 0.5$), Fig. 7. The values of the height of the energy barrier of pore formation for different lateral tensions and projection of the boundary director are presented in Table I.

Table I. The height of the energy barrier of the pore formation process calculated for different lateral tensions $\sigma$ and projections of the boundary director $|n_0|$.

| | | Director projection, $|n_0|$ | | | |
|---|---|---|---|---|---|
| | | **Lipidic pore** | **0.3** | **0.4** | **0.5** |
| **Lateral tension, $\sigma$** | **0.1 mN/m** | 50.5 $k_BT$ | 37.7 $k_BT$ | 36.4 $k_BT$ | 33.5 $k_BT$ |
| | **4 mN/m** | 60.3 $k_BT$ | 46.1 $k_BT$ | 39.2 $k_BT$ | 31.8 $k_BT$ |

According to the Derjaguin-Gutop model, lateral tension should decrease the energy barrier of pore formation. This model assumes that the line tension (the energy of the pore edge divided by its perimeter) is independent on the pore radius and the membrane lateral tension. However, the line tension does depend on both the radius and lateral tension [61]. The line tension is thought to be determined by the energy of elastic deformations arising at the pore edge. This energy increases monotonically with increasing lateral tension. In other words, deformations require more energy when the membrane is under lateral tension [61,62]. When the pore radius is large enough, the term ($-\pi R_0^2 \sigma$) makes the dependence $W(R_0)$ go down. At small radii, when the area of the pore is small, as well as the term ($-\pi R_0^2 \sigma$), the growing lateral tension increases the line tension thus causing the total pore energy to increase [Fig. 3(a)]. At some radius, the energy increase due line tension growing is approximately compensated by the energy decrease provided by the term ($-\pi R_0^2 \sigma$) [$R_0 \approx 2.3$ nm, Fig. 3(a)]. As a result, the energy at this pore radius is almost independent on the lateral tension. As the radius of the pore lined with two peptide molecules is small enough (i.e., $R_p = 1.6$ nm $< 2.3$ nm), the energy should generally grow up with increasing lateral tension. This is the case for the purely lipidic pore (see Table I). However, undeformable inserted peptides substitute deformable lipidic material and partially relax the deformations by introducing a curvature that fits the pore edge. This allows compensating the energetically unfavorable consequence of the lateral tension growing, especially, when the director projection at the peptide boundary is large (Table I).

## IV. DISCUSSION

### A. Non-local mechanism of pore formation by amphipathic peptides in high concentration

Various AMPs induce through pores in bacterial membranes thus leading to the microbial cell death. The pore formation process can occur by two principal mechanisms, local and non-local. According to the non-local mechanism, AMPs at high concentration rapidly insert into the external monolayer of a closed membrane of bacterial cell or, e.g., GUV, and induce a high lateral pressure in this monolayer and a lateral tension in the opposing monolayer. Here for the first time (as far as we know) we analyzed the energy landscape of pore formation in such asymmetric conditions. We demonstrated that applied lateral pressure/tension can significantly decrease the energy barrier of pore formation [Fig. 3(b)]. Pressure/tension is a non-local parameter, as it is the same in the whole membrane. Thus, the energy barrier decreases non-locally, i.e., pore formation is facilitated at any point of the closed membrane.

Recently, the carpet model has been proposed to describe the action mechanism of amphipathic peptides applied in very high concentrations [63]. This model assumes a detergent-like action of amphipathic peptides. Formally, typical detergents are amphipathic. When applied in concentration, detergents disrupt a membrane into micelles containing lipids of the membrane and molecules of the detergent. In this case, the membrane becomes fragmented, and the area of the lipid bilayer decreases due to (partial) dissolution. The carpet model suggests a similar mechanism of membrane disruption by amphipathic peptides applied in very high concentrations [63]. However, in the experiments described in the companion paper [38], the area of the GUV membrane increased rather than decreased upon peptide adsorption in high concentrations (see Table V of the companion paper [38]). This means that the membrane dissolution threshold was not reached and the conditions for the carpet mechanism of membrane disruption did not occur. In the regime of our non-local mechanism, the membrane is not dissolved or fragmented either. The peptide-induced increase in the area of the outer monolayer of a closed membrane (e.g., GUV) is considered to be the reason of lateral pressure/tension generation leading to reversible pore formation, but not to irreversible membrane fragmentation or dissolution as supposed in the carpet model.

We considered the range of lateral pressures/tensions, which are readily achievable in a physical experiment. In [4], the interaction of the amphipathic peptide PGLa with GUV membrane composed of DOPC and dioleoylphosphatidylglycerol (DOPG) has been investigated. It is shown that the membrane area increases up to about $\delta = 3$ % upon PGLa adsorption onto the external monolayer of a GUV membrane. The arising lateral pressure/tension can be estimated as $|\sigma_0| = k_a \delta \approx 133 \times 0.03 \approx 4$ mN/m. Experiments show that such an increase in the membrane area leads to an intensive translocation of PGLa molecules to the inner monolayer of the GUV, which means that multiple pores are formed [4]. A similar increase in the area of the membrane we observed in our experiments with melittin and magainin-1 (see Table V of the companion paper [38]) applied in high bulk concentration. Therefore, this effect reflects a common mechanism regardless of the particular chemical structure of the AMP.

In the case of asymmetric pressure/tension, the formation of metastable pores with large radius and long lifetime is predicted [Fig. 3(b)]. We relate the electric activity of such pores to spike electric signals observed in the companion paper [38]: they have relatively high amplitude and last about milliseconds to tens of milliseconds (Figs. 2, 3 of the companion paper [38]). Metastable pores are hydrophilic [see vertical arrows in Fig. 3(b)], and in this state membrane monolayers subjected to lateral pressure (upper) and lateral tension (lower) meet each other and become continuous. This should result in a lipid flow from the upper to the lower monolayer, which is substantially a non-equilibrium dissipative process that cannot be formally considered in the framework of our theoretical model. However, in such a flow, lipid molecules in the upper and lower monolayers should move in opposite directions: towards and outwards the pore edge, respectively. Due to principal difference in the molecule velocities, this should unavoidably lead to large forces of inter-monolayer friction, which is shown to be about an order of magnitude higher compared to in-monolayer friction [64]. For this reason, the lipid flow between the monolayers can be considered slow, meaning that the membrane is under quasi-equilibrium conditions. Within such limitations, we calculated the dependences $W(R_0)$ presented in Fig. 3(b) beyond the vertical arrows, i.e., for large radii, where hydrophilic pores are already formed and membrane leaflets become continuous.

Hypothetically, the lipid flow can induce an additional expansion of the pore as compared to the pore evolution in a symmetric membrane without the flow. The size of the pore area is several square nanometers, much smaller than the membrane area of a typical GUV. Therefore, the latter acts as a reservoir for lipid rearrangement around the pore. The flow can be considered stationary, with a characteristic establishment time of about $\tau = \rho R_0^2 / \eta$ ($\rho$ is the characteristic density and $\eta$ is the characteristic viscosity of the lipid-water environment of the pore) [65]. Considering the maximum $\rho$ and the minimum $\eta$, i.e., for water, we obtained an upper estimate for $\tau \sim 10^{-12}$ s. In [65] the expansion of a hydrophilic pore in a planar bilayer, where the monolayers are under unequal tensions, was studied. The analysis takes into account both intra-and extra-monolayer friction, and the line tension $\gamma$ of the pore edge. The study shows that the pore expansion rate is determined by the sum of the tensions of the two monolayers, which is approximately zero in our asymmetric case, and thus has no effect on the pore expansion. The physical meaning of this result is the following: as the pore expands, the same amount of lipid material from each monolayer flows into the reservoir and, as their tensions have opposite signs, the total work on the system is zero.

The lipid flow through the edges of the pores gradually relaxes the peptide-induced difference in the lateral pressure/tension between the monolayers. The relaxation should lead to a closure of metastable pores as the energy in the corresponding local minimum grows [Fig. 3(b)], compare minima of green and gray curves]. However, the relaxed lateral pressure in the external monolayer facilitates further adsorption of AMPs from the water solution onto the external monolayer [62]. The adsorption generates a difference in the lateral pressure/tension between two monolayers once more, metastable pores open, an arising lipid flow relaxes the pressure/tension, and so on. This cyclic opening of large metastable pores can last till the surface concentrations of AMP in the external monolayer become so high that the probability of collision of metastable pores with diffusing peptide molecules approach unity within the average lifetime of the pore. Upon the collision, the pore edge can capture a peptide molecule, as in such a configuration the elastic energy of the system has (at least) a local minimum [66] (Fig. 7). Further, the AMP molecule can escape from the pore edge either to the same or to the opposite lipid monolayer, or the pore edge can collide with another AMP molecule and capture it as well. Such capturing is accompanied by an increase in the pore radius, i.e., its electric conductance increases, and so does the average lifetime of the pore. We relate the electric activity of the pores, which were captured peptide molecules to their edges within their lifetime, to the multi-level and square-top signals observed experimentally in the companion paper [38] (see Fig. 2 of the companion paper [38]). As peptide molecules directly participate in the formation of the edges of such pores, we refer to the mechanism of such pore formation as the local one. As the edges of the pores are formed by both lipid and peptide molecules, such pores correspond to toroidal ones [11]. The translocation of AMP molecules through the pore edges should lead to a gradual equilibration of AMP concentrations at the external and internal monolayers of the membrane. After reaching the equilibrium, the formation of large pores by the non-local mechanism should stop in GUVs and become suppressed in patch-clamp conditions. This reasoning agrees with experimental observations of the companion paper [38] (see Figs. 1, 4, 5 of the companion paper [38]).

In the companion paper [38], pore formation induced by magainin or melittin was studied experimentally. Experiments were performed on GUVs and planar lipid bilayers; pore formation was registered as calcein flow and the appearance of electrical conduction, respectively. In the case of GUVs, the mechanism of pores formation through which calcein molecules (~1 nm in size) can pass follows directly from our theoretical model: amphipathic peptides applied in high concentration adsorb onto the closed GUV membrane and induce sufficiently high lateral pressure in the outer monolayer and lateral tension in the inner monolayer. This leads to a decrease in the energy barrier of pore formation and an increase in the energy barrier of closure of large metastable pores [Fig. 3(b)]. In the case of the electrophysiological experiment [38] on planar lipid bilayers, the application of our model of non-local pore formation is less obvious, as a planar lipid bilayer is connected to a large lipid reservoir that maintains a constant lateral tension in both monolayers of the membrane. However, the patch-clamp technique was used in this experiment [38]. Amphipathic peptides put inside a glass micropipette adhered to the upper monolayer of the planar lipid bilayer. The glass micropipette thus isolated a small patch of the upper lipid monolayer. Adsorption of amphipathic peptides from the micropipette onto the isolated patch of the upper monolayer should induce a lateral pressure in this patch, while the opposing monolayer remained continuously connected to the lipid reservoir and thus remained subjected to the constant lateral tension of about 1 mN/m (Fig. 8) [46].

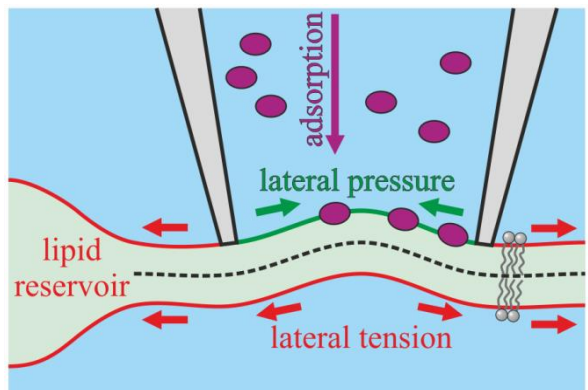

FIG. 8. The scheme of the lateral pressure/tension generation under conditions of the patch-clamp experiment. The micropipette (shown in gray color) with AMP solution (shown as purple ellipses) isolates the patch of the upper monolayer. The whole bilayer is subjected to lateral tension (red arrows) imposed by the lipid reservoir. Peptide adsorption to the isolated patch of the upper monolayer generates lateral pressure (green arrows) there. An asymmetric lateral pressure/tension appears under the micropipette.

In such a configuration, an asymmetric lateral pressure/tension appears, and the membrane patch in the glass micropipette can be qualitatively described by our theoretical model, although most probably in this case the total tension ($\sigma_u + \sigma_l$) may deviate from zero. Experimentally, the formation of pores was observed as the appearance of spikes of electric conductance (see Fig. 2 of the companion paper [38]). We attributed the spike conductance to the formation of metastable pores of radius of one to several nanometers according to our model. The experimentally determined average lifetime of such pores is of the order of milliseconds; thus, they are relatively long-living compared to the metastable pores in membranes with symmetrically applied lateral tension, whose lifetime is estimated to be of the order of nanoseconds from molecular dynamics modeling [25]. Pore formation allows the lipid to flow from the upper to the lower monolayer and relaxation of the lateral pressure/tension. This leads to pore closure. However, amphipathic peptides from the micropipette interior continue to adsorb onto the isolated patch of the upper monolayer; the adsorption leads to an increase in the lateral pressure; once more the bilayer patch becomes asymmetric with respect to the lateral pressure/tension; metastable pores appear, and so on.

The non-local mechanism of pore formation in closed (or isolated) membranes by amphipathic peptides should be weakly sensitive to a particular chemical structure of the AMP molecule. The adsorbing AMP should introduce some area to the external monolayer of the closed membrane, and the affinity of the AMP to the membrane should ensure the sufficient surface concentration (and area increase) to achieve the necessary value of the lateral pressure/tension. Most amphipathic molecules can readily do this. As AMP molecules by the non-local mechanism do not directly participate in the formation of the pore edge, their lateral area, depth of insertion into the lipid monolayer, particular sequence of amino acids, etc., are not important. In other words, the non-local mechanism of pore formation is very non-specific. Thus, our model allows estimating the extent to which the selectivity and specificity of the pore-forming activity of amphipathic peptides can be principally improved. This mechanism works only when the peptides are applied asymmetrically to the membrane; it provides the formation of purely lipidic pores. When the peptides are applied symmetrically, this mechanism is irrelevant, and no lipidic pores are observed in this case: the edges of the pores formed in symmetric conditions are always occupied by AMP molecules [11,15].

In a series of works [31–34] reviewed in [35], kinetic schemes were proposed and experimentally proved for several AMPs. It was shown that amphipathic peptides can make LUV membrane permeable for water-soluble fluorescent dye by two mechanisms: all-or-none (either complete or no leakage of the water-soluble dye from LUVs) and graded (continuous distribution of the concentration of the water-soluble dye in LUVs). The all-or-none mechanism proposed by Gregory et al. in [31] has some points that are similar to our non-local mechanism. The main difference is that the "all-or-none" model is qualitative and speculative to some extent; it is not based on a solid physical theory. E.g., Gregory et al. write: "The bound peptides probably cause bilayer thinning and positive curvature strain because of the excess mass on the outer monolayer of the membrane. A point of rupture is eventually reached where the bilayer yields, allowing the contents to leak out all at once". The formulated all-or-none model does not address how exactly and why the "bilayer thinning" promotes pore formation; how exactly and why the "positive curvature strain" in only the outer monolayer promotes pore formation. The "excess mass on the outer monolayer" does not necessarily mean the appearance of the "positive curvature strain" leading to pore formation; the contrary instance is an electrostatic binding of peripheral proteins to membrane [64]. In addition, a specific "point of rupture" does not exist for membranes: the formation of a pore is a stochastic process, which is characterized by an average waiting time [26]. In this view, there is no specific "point of rupture" determined by a particular set of numerical values of the system parameters; there is only a dependence of the average waiting time of pore formation on these parameters. The waiting time is, in turn, determined by the energy barrier of the corresponding process. The "bilayer thinning" and "positive curvature strain" are local deformations: they may be induced only in the nearest vicinity (about several nanometers) of the incorporated peptide molecule. In this sense, the all-or-none model formulated by Gregory et al. [31] does not exactly correspond to our non-local mechanism. On the contrary, we considered the asymmetry in the lateral pressure/tension across two lipid monolayers as the main driving force of pore formation. Lateral pressure and tension are definitely non-local parameters. We related quantitatively the surface concentration of incorporated peptides to the induced asymmetry in the lateral pressure/tension from the experimental results on membrane area increase upon amphipathic peptide adsorption obtained in the companion paper [38] and in [4]. Further, we related quantitatively the asymmetry in the lateral pressure/tension to the energy barrier of pore formation [Fig. 3(b)], which provides the prediction on the average waiting time of pore formation. The calculations were performed for a particular lipid (DOPC) and for particular peptides (magainin and melittin). However, the developed model is quite general: substituting the elastic parameters of another lipid and geometric parameters of another peptide would provide the quantitative prediction on the average waiting time of pore formation for any pair of lipid and peptide. The all-or-none model described in [31] cannot quantitatively predict anything, as it is based on pure phenomenology: the results allow only to conclude, that interaction of a particular peptide (cecropin A) with lipid membranes of particular compositions (palmitoyloleoylphosphatidylcholine, POPC, and palmitoyloleoylphosphatidylglycerol, POPG, from pure POPC to POPC:POPG 50:50) at particular concentration (1 μM cecropin A) yields pores within determined kinetics. However, these results do not clarify the underlying physics of the process on mechanical level, and thus do not have the predictive power: the experiments should be repeated for each new pair of lipid and peptide and new values of concentrations.

The calculations performed in the framework of the proposed non-local mechanism of lipid pore formation predicted the drop of the energy barrier of metastable pore appearance from 46.5 $k_BT$ at almost zero lateral tension to 32.2 $k_BT$ at asymmetric lateral tension/pressure of 4.5 mN/m. The waiting time to overcome the energy barrier of 46.5 $k_BT$ may be estimated as about 2 days. In the companion paper [38] it was observed that the addition of amphipathic peptides in pore-forming concentrations leads to an increase in GUV diameter by about 2%, which corresponds to about a 4% increase in its area. This increase should lead to the appearance of lateral tension/pressure of about $k_a\delta = 133$ mN/m × 0.04 ≈ 5.3 mN/m. Linearly extrapolating the dependence of the theoretically predicted energy barriers of metastable pore formation according to the non-local mechanism we obtain that the energy barrier is 28.6 $k_BT$ at lateral tension/pressure of 5.3 mN/m. According to theoretical estimations, the waiting time to overcome this energy barrier is of the order of 3 ms, i.e., the limiting stage in this process is most probable the peptide adsorption. This quantitative prediction could be experimentally disproved if amphipathic peptides applied in the concentration that led to increase in GUV diameter by 2% did not induce calcein leakage. Additionally, the non-local model could be disproved if amphipathic peptides applied in low concentration yielding no increase in GUV diameter, had induced an appreciable calcein leakage.

## B. Local mechanism of pre-pore formation by amphipathic peptides in low concentration

When applied in low concentration, amphipathic peptides cannot generate a sufficient difference in lateral pressure/tension in two membrane monolayers to induce pores by the non-local mechanism. In the framework of Derjaguin-Gutop theory, this means that AMP cannot decrease the energy barrier of pore formation by modifying the lateral tension term of pore energy. However, the energy barrier $\Delta E = \pi\gamma^2/\sigma$ can still be diminished by decreasing the line tension of the pore edge $\gamma$. To do so, AMPs have to directly participate in the formation of the pore edge by partial substituting lipid molecules. We refer to the corresponding mechanism of pore formation as the local one, as in this case pores can form only locally, in the immediate vicinity of peptide molecules. As the edges of such the pores comprise both lipid and peptide molecules, these pores correspond to toroidal ones [11].

Starting from the ground state, we elaborated a continuous trajectory of pore formation in the middle between two parallel peptide molecules standing in the registered configuration. This region of the membrane accumulates most elastic stress induced by peptides [Fig. 4(a)]; thus, the formation of a membrane defect is most probable to occur there. The registered configuration of two straight parallel peptide molecules corresponds to the global minimum of the membrane elastic energy; this configuration is the ground state of the system. When a circular hydrophobic defect is formed in between the peptides, it becomes energetically favorable to bend the peptides into a ring-like structure (Fig. 5). We assumed that two halves of the ring-like structure, each formed by a bent peptide molecule, do not shift laterally. However, it is not actually a hypothesis, as such a configuration also corresponds to the global energy minimum of the system at a fixed radius of the hydrophobic defect. From our calculations it follows (data not shown) that when a hydrophobic defect of some radius is formed in between the peptides, the global minimum of the energy (at fixed radius of the hydrophobic defect) corresponds to the bent peptides located at approximately the same distance $L$ between their centers (see the inset in Fig. 5). This means that when the hydrophobic defect forms, two straight parallel peptides bend to a ring-like structure, but they should not shift laterally. Further, the trajectory of pore formation can be described by the tilt $n_t$ of the peptide ring [Fig. 6(b)]. The configuration where the two peptides ring is located at the pore equator corresponds to the terminal point of the continuous trajectory; at this point $n_t = 1$. At this terminal point, the total energy of the system has a local minimum (Fig. 7), i.e., the small hydrophilic pore with two peptide molecules located at its equator can be metastable with a finite lifetime. We refer to this configuration as the $H^+$-pore. In this configuration, the $H^+$-pore radius measured at the lipid monolayer neutral surface is equal to $R_p = 2L_p/(2\pi) \approx 1.6$ nm. Thus, the $H^+$-pore lumen radius should be smaller by the size of the polar part of the lipid molecule and it additionally depends on the depth the peptide is inserted into the lipid monolayer. The local minimum depth of the system total energy increases when the projection of the boundary director $|n_0|$ increases, and decreases when lateral tension increases (Fig. 7). The largest projection of the boundary director yields the lowest energy barrier of the $H^+$-pore formation. Such a projection is achieved when the amphipathic peptides are shallowly inserted into the lipid monolayer, and, thus, under optimal conditions the pore lumen radius should be rather small, e.g., about $R_p - 2r_0 \approx 0.3$ nm, meaning that the pore should be impermeable to large substances like water-soluble fluorescent dyes and solvated ions, but permeable to protons. The formation of the pores impermeable to calcein, but permeable to protons was observed experimentally in the companion paper (see Figs. 4, 5, 7 of the companion paper [38]). For large enough $|n_0|$, the energy barrier of pore formation decreases when lateral tension increases, i.e., for a large enough projection of the boundary director, the process of the $H^+$-pore formation is stretch-activated, which agrees with the experimental observations in [18] for magainin. We can thus infer that for magainin it should be: $|n_0| \geq 0.5$ (see Table I), which also agrees with the prediction $|n_0| = 0.48$ in [60]. Peptides can escape from $H^+$-pore equator either onto the initial or onto the opposite side of the membrane with equal probability. This provides a theoretical mechanism of peptide translocation through the membrane when large pores observable in experiments on dye leakage or in measuring the $K^+$ electric current through the membrane do not form.

According to the theoretical consideration, translocated peptides are not inactivated: they can also form registered configurations [Fig. 4(a)] using as partners the peptides from their own lipid monolayer or from the opposing lipid monolayer; in the latter case, the depth of the energy minimum corresponding to the registered configuration is even greater [Fig. 4(b)]. This is in line with our experimental results [38] showing the protective effect of melittin and magainin added from the GUV interior against pore formation (see Table IV of the companion paper [38]). Registered configurations can further lead to the formation of metastable $H^+$-pores in accordance with the above-described mechanism. Thus, some equilibrium population of metastable $H^+$-pore in the membrane is predicted to exist, via which the inner and outer lipid monolayers can exchange by lipids and peptides.

When the surface concentration of peptides adsorbed onto the membrane is high enough, an additional peptide molecule can meet the metastable $H^+$-pore formed by two peptide molecules during its lifetime. This additional peptide molecule can be captured by the $H^+$-pore edge, and thus a metastable pore of increased radius is formed. Further, this pore can either expel the peptide molecule from its edge or meet and capture an additional peptide molecule; and so on. The evolution of the average radius of the ensemble of such metastable pores ($H^+$-pores) should depend on the average lifetime of the $H^+$-pores, the surface concentration of the peptides, and their diffusion coefficients. We may thus conclude that relatively large lipid-peptide (toroidal) pores can be formed predominantly by the local mechanism, when peptides participate in the pore edge formation. This mechanism can explain the appearance of large pores when AMPs are added symmetrically, to both sides of the membrane, in a high bulk concentration [9,11,15].

In our calculations, we assumed that the peptide molecules lie in the equatorial plane of the small $H^+$-pore parallel to the plane of the unperturbed membrane. Formally, the peptides can tilt out of this orientation [5,67,68]. However, accounting for this tilt would lead to a loss of the rotational symmetry of the system, thus greatly complicating the calculations. Assuming that the peptides orientation at the $H^+$-pore equator is fixed, we excluded a possible degree of freedom of the system. Therefore, the calculated energies and energy barriers should be formally considered as an upper estimate. Thus, we predicted that just two AMP molecules should be sufficient to form a small $H^+$-pore impermeable to water-soluble fluorescent dyes and solvated ions, meaning that such small $H^+$-pore can be formed even at very low surface concentrations of peptides.

The energy barriers of pore formation by the local mechanism are of the order of 33–38 $k_BT$ at a low lateral tension (Table I). Are these energy barriers high or low? For membrane fusion, it is estimated that the energy barrier of the height of 40 $k_BT$ can be surmounted within the average waiting time of about 1 min. at the expense of the energy of thermal fluctuations [69]. The energy of the stalk intermediate calculated in this work is about 37 $k_BT$. In [70] the minimal energy of stalk of the optimal shape formed between two DOPC bilayers is 43 $k_BT$. In the paper on membrane fusion [71], the membranes, for which the energy barrier of stalk formation was estimated as 40 $k_BT$, fuse readily in the experiment, while the membranes made from DOPC (with the estimated energy barrier of stalk formation of 44.5 $k_BT$) do not fuse within tens of minutes. Thus, we may conclude that the estimation of 40 $k_BT$ obtained in [69] for energy barriers of probable spontaneous membrane processes is quite accurate. In the estimation of the average waiting time of stalk formation (~1 min. for the energy barrier of 40 $k_BT$) the pre-exponential factor may play a role. In the work [69], the waiting time of fusion stalk formation is estimated as $\tau \sim \dfrac{1}{\omega S N} \exp\left[\dfrac{\Delta W}{k_B T}\right]$, where $S \sim 100$ nm$^2$ is the area of the stalk (fusion site), $N \approx$ 100 is the number of stalks (fusion sites) in the system. Applying similar Arrhenius-like approach for the estimation for the average waiting time of pore formation, one should set $S \sim 1$ nm$^2$ (a typical area of the pore) and $N$ equal (by the order of magnitude) to the number of lipid molecules in the membrane; $N$ should be close to the total area of the membrane expressed in nm$^2$, as the area per lipid is about 1 nm$^2$. For the membrane of spherical GUV of 1 μm diameter, this estimation yields $\tau \sim 1.5$ s for $\Delta W = 40\ k_BT$. The details of the pre-exponent factor (e.g., the particular area of the pore, the particular area of the membrane, etc.) have minor influence on the waiting time value compared to the height of the energy barrier $\Delta W$, as the latter stands in the exponent. The diffusion coefficient of the membrane-inserted amphipathic peptide is of the order of 1 μm$^2$/s, i.e., within 1 min. each peptide molecule runs about 15 μm over the membrane surface. Thus, even at very low surface concentrations, the probability of collision of two peptide molecules is quite high. In Table I, the energy barrier of lipidic pore formations of radius $R_0 =$ 1.6 nm at low lateral tension is about 50 $k_BT$. As 50 $k_BT > 40\ k_BT$, we can conclude that it is very unlikely to observe experimentally such pores to form spontaneously, which agrees well with the experimental data in [38]. For boundary director projections corresponding to melittin ($n_0 = 0.4$) and magainin ($n_0 = 0.5$) the energy barriers are 36.4 $k_BT$ and 33.5 $k_BT$, respectively. Both values are less than 40 $k_BT$ meaning that pore formation is very probable within the waiting time of the order of 1 min.

Thus, the developed continuous trajectory of pore formation by two molecules of amphipathic peptide is energetically possible, i.e., the energy barrier on the trajectory is less than 40 $k_BT$. However, the probability of pore formation (in the membrane patch of unit area within unit time) should also depend on dynamical quantities such as the lipid or peptide diffusion coefficients, rotational diffusion of the peptides, average surface concentration of the peptides, etc. At a low surface concentration of peptides, weakly cooperative configurations, from which the pores can be formed in the system, may occur extremely rare despite a relatively low energy barrier of the process.

We estimated theoretically that the energy barrier of $H^+$-pore formation by two molecules of magainin (the boundary director values $|n_0| = 0.5$) at a low lateral tension ($\sigma = 0.1$ mN/m) is 33.5 $k_B T$ (see Table I). Thus, the characteristic timescale of the $H^+$-pore formation can be roughly estimated as $\tau = \tau_0 e^{33.5 - 40} \approx \tau_0 e^{-6.5} \approx 90$ ms, where $\tau_0 = 1$ min. The rough estimation of the characteristic lifetime of the $H^+$-pore yields: $\tau = \tau_0 e^{(39.5 - 21.8) - 40)} \approx \tau_0 e^{-22.3} \approx 0.01$ μs. These rough estimations are based on the assumption that the characteristic frequency standing in the pre-exponential factors weakly depend on the particular membrane process (membrane hemi-fusion vs. $H^+$-pore formation and $H^+$-pore closure), and are mainly determined by the characteristic frequency of lipid thermal fluctuations. However, this assumption is not strictly valid for the process of the $H^+$-pore closure, as the non-ideal diffusion of two interacting peptide molecules should be taken into account, as its characteristic time is several orders of magnitude larger. The coefficient of the ideal diffusion of a membrane-adsorbed peptide is rather low, of the order of 1 μm$^2$/s, and it seems reasonable that it is the non-ideal diffusion that mainly determines the lifetime of the $H^+$-pore. Moreover, in the processes of formation and closure of the $H^+$-pore, the peptide molecule moves either towards or outwards the pore edge (equator). This should lead to a motion of the lipids in the two opposing membrane leaflets in opposite directions. E.g., when the $H^+$-pore is formed by the peptide molecule located in the upper monolayer, the lipid of the upper monolayer would move towards the edge of the $H^+$-pore, while the lipids of the lower monolayer would move from the $H^+$-pore edge [compare Fig. 6(b) and Fig. 6(c)]. Due to the principal difference in molecule velocities, this should unavoidably lead to large forces of inter-monolayer friction, which is shown to be about an order of magnitude higher compared to the in-monolayer friction [64]. This friction can be an additional factor for much slower formation and closure of $H^+$-pores as compared to the rough estimation based on the results of the work [69], as presented above.

Our local mechanism does not exactly correspond to the graded mechanism of fluorescent dye release from LUVs described in, e.g. [33]. The minimal pore formed by two peptide molecules according to the local mechanism is impermeable to calcein and $K^+$; thus, it cannot provide the graded release of calcein. Within its lifetime, the minimal pore can capture an additional molecule of the peptide and become larger, but this possibility was discussed only qualitatively; the corresponding process and statistics were not quantitatively analyzed here.

## C. Protective effect of membrane pre-treatment by AMP in low concentration

According to the model of the local mechanism of $H^+$-pore formation, a pre-treatment of a planar lipid bilayer with amphipathic peptides in low concentration yields a population of small metastable $H^+$-pores representing multiple continuous connections between the upper and lower monolayers of the membrane. Through such connection, the monolayers can exchange lipids and peptides. Generally, AMP adsorption to the closed membrane generates lateral pressure in the outer leaflet and lateral tension in the inner leaflets, which eventually causes large metastable pores to form in accordance with the non-local mechanism. However, when the membrane was pre-treated by a low bulk concentration of AMP, an intensive adsorption of subsequently added AMP in a high bulk concentration would induce a lipid and peptide flow to the inner leaflet through the small metastable $H^+$-pores instead of generation of large lateral pressure/tension gradients. Thus, such small $H^+$-pores can smooth otherwise sharp gradients of the lateral pressure/tension protecting membranes from the formation of large leaky pores. In this sense, the pre-treatment of a membrane with amphipathic peptides in a low bulk concentration protects the membrane against a rupture upon subsequent application of amphipathic peptides in a high bulk concentration. In this case, an intensive adsorption of the AMP to the outer membrane leaflet only induces a flow of lipids and peptides from the outer leaflet to the inner leaflet of the membrane via small $H^+$-pores, thus resulting in an about zero lateral pressure/tension in both membrane leaflets. This theoretical scheme explains the protective effect of low bulk concentrations of the AMP.

In the companion experimental paper [38], it is shown that upon the application of AMP (magainin and melittin) in nanomolar bulk concentrations to GUVs, calcein leakage does not occur or is highly suppressed (see Figs. 4, 5, 8 of the companion paper [38]), while GUVs become leaky for protons (see Figs. 7, 9 of the companion paper [38]). Similar membrane leakage for protons was observed for melittin and arenicin-2, but only in micromolar bulk concentrations [72]. Application of AMP in a high bulk concentration leads to a fast leakage of fluorescent dye from the GUV interior (see Figs. 4, 5, 8 of the companion paper [38]) and yields relatively large long-living pores in electrophysiological experiments (see Figs. 1, 3 of the companion paper [38]). We may thus suppose that small $H^+$-pores formed in response to AMP applied in a low bulk concentration do not allow calcein or even solvated ions to pass through, i.e., their lumen diameter should be smaller than ~1 nm. The application of AMP in a high bulk concentration leads to the formation of ion-conducting and calcein-permeable pores. In our experimental setup, we did not apply a background lateral tension ~0.5 mN/m, in contrast to works [4,17]. Because of the absence of the lateral tension in our experiments, GUVs are not completely destroyed upon application of AMP in a high bulk concentration {see Figs. 4(a), 5(a) of the companion paper [38]}. When a high bulk concentration of AMP is applied after the pre-treatment of membranes with a low concentration of AMP, the formation of large pores is greatly or completely suppressed as compared to the case of immediate treatment of membranes with AMP in a high bulk concentration (see Tables I, II, IV of the companion paper [38]).

The experiments in the companion paper were performed for magainin and melittin. It was shown that a low bulk concentration of magainin protected completely the membrane against the formation of large pores upon subsequent application of magainin in a high bulk concentration (see Table IV of the companion paper [38]). However, the protective effect of melittin in low concentration was not complete: the number of large pores formed after the application of melittin in a high bulk concentration decreased, but never became zero (see Table IV of the companion paper [38]). In our elastic approach, the peptide is characterized by its diameter, length and the projection of the boundary director $|n_0|$. Among these parameters, $n_0$ is predicted to be the most effective regulator of the pore formation process. About 1/3 of the melittin molecule is relatively deeply inserted into a lipid monolayer, in contrast to magainin, which is inserted shallowly. Thus, in average, the projection of the boundary director $|n_0|$ should be smaller for melittin than for magainin. This conclusion agrees with the results of the work ref. [60]. In the present paper, we considered the pore formation process for different values of $|n_0|$. The difference between the energy in the states with $n_t = 1$ and $n_t = 0$ determines the average number of small metastable $H^+$-pores with AMP at the pore equator (e.g., in accordance with the Boltzmann distribution); this number should decrease with decreasing $|n_0|$ [Fig. 7(a)]. This means that melittin should form a fewer number of $H^+$-pores, and their lifetime should be shorter. The small equilibrium population of melittin-induced short-living $H^+$-pores may simply not be able to effectively relax the large difference in the lateral pressure/tension in the outer and inner lipid monolayers arising upon applying melittin in a high bulk concentration; this explains the incomplete protective effect of low concentrations of melittin (see Table IV of the companion paper [38]).

Qualitative consideration in the framework of the local mechanism allows us to conclude that the formation of equilibrium population of metastable $H^+$-pores formed by just two peptide molecules is possible from the energetic point of view: the corresponding energy barrier is about 33.5 $k_BT$ for magainin. This population of the $H^+$-pores is predicted to at least partially protect the membranes from an intensive formation of large long-living pores upon addition of amphipathic peptides in high concentration. Experimentally, it potentially could be observed that the kinetics of calcein leakage is independent on the pre-adsorption of amphipathic peptides in low concentration; such an observation could disprove the local model.

The membrane protection by a pre-treatment with a low concentration of AMP can lead to a negative side-effect of using amphipathic peptides as antimicrobial agents. When applied in a low concentration, AMPs would protect bacterial membranes from destruction. In addition, if adsorbed to plasma membranes of normal host cells in low concentration, AMPs can potentially relax the asymmetry of membranes with respect to the lipid composition of those lipid monolayers. This can lead to the exposure of negatively charged lipids at the outer membrane leaflet, while normally negatively charged lipids are located at the inner leaflet of plasma membranes of mammalian cells [73,74]. Such exposure can potentially lead to the immune system misclassifying a normal cell as a pathogen, resulting in an inflammation.

**D. Assumption of the elastic theory**

In the present work, we calculated the elastic energy of membrane in the framework of continuum theory of elasticity. The sizes of the pores formed in the membrane are of the order of units of nanometers, i.e., they are comparable with molecular dimensions. Formally speaking, no continuum theory can be applied on the molecular scales. Moreover, in our theoretical considerations, we used the linear theory of elasticity (applicable formally for small deformations only) to describe highly deformed structures. The applicability limit of a continuum linear theory can be determined only from an external (with respect to the theory) reasoning, e.g., based on the data from physical experiment or molecular dynamics. For example, consider a nucleus of a heavy atom, like uranium. The nucleus is definitely a non-classical object that should be analyzed substantially in the framework of quantum mechanics. However, there is a well-known liquid-drop model of heavy nuclei [75]. This model treats the nucleus as a drop of incompressible fluid characterized by substantially classical (and even continuum) parameters like surface tension, Coulomb interaction of particles in the fluid, etc. For heavy atoms, the basic equation of the model, Weizsäcker formula, provides results that are in quantitative agreement with experimental data. Obviously, no classical and especially continuous theories can be used to consider the energy of atomic nuclei. The assumptions on which the Weizsäcker formula is based are obviously invalid. Nevertheless, experiments prove that the limits of applicability of the Weizsäcker formula are far beyond the limits of formal validity of its basic assumptions. We stated that the applicability limit of a continuum linear theory can be determined only from an external (with respect to the theory) reasoning, e.g., based on the data from physical experiment or molecular dynamics, meaning this sense. In [46], the experimental data and molecular dynamics modeling are related to calculations within the linear theory of elasticity for highly deformed membrane structures. It is shown that the linear theory of elasticity is quite accurate even when applied on molecular scales to describe highly deformed lipid monolayers. Molecular dynamics can definitely be used to test predictions made based on continuum physics, as the lateral dimensions of the modeled membranes are much larger than characteristic lengths of deformations, and the length of the obtained trajectories is much larger than the characteristic time of thermal fluctuations.

The model of pore formation, very similar to models used here, has been developed earlier in [25]. This model is also based on the continuum linear theory of elasticity. In addition, molecular dynamics simulations of spontaneous pore closure have been performed in this work. The continuous trajectory of pore formation (and closure) calculated within the developed model based on continuum linear theory of elasticity agrees quantitatively with the molecular dynamics data for three different lipids. Thus, although it is generally incorrect to use continuum linear theories of elasticity to describe molecular-scale deformations of large amplitude, for our particular model, the applicability limit determined from molecular dynamics does allow the process of pore formation to be described.

## ACKNOWLEDGMENTS

This research was funded by the Ministry of Science and Higher Education of the Russian Federation.

## DATA AVAILABILITY STATEMENT

The data that support the findings of this study are available from the corresponding author upon reasonable request.

## AUTHOR DECLARATIONS

### Conflict of Interest

The authors have no conflicts to disclose.

# Dialectics of antimicrobial peptides II: Theoretical models of pore formation and membrane protection

Oleg V. Kondrashov[1], Marta V. Volovik[1], Zaret G. Denieva[1], Polina K. Gifer[1], Timur R. Galimzyanov[1, #], Peter I. Kuzmin[1], Oleg V. Batishchev[1], Sergey A. Akimov[1]

[1]Laboratory of Bioelectrochemistry, A.N. Frumkin Institute of Physical Chemistry and Electrochemistry, Russian Academy of Sciences, 31/4 Leninskiy prospekt, Moscow, 119071, Russia

#Present address: JetBrains, JetBrains Research, 23 Christoph-Rapparini-Bogen, München, 80639, Germany

Correspondence: olegbati@gmail.com (O.V.B.); akimov_sergey@mail.ru (S.A.A.)

## THEORETICAL BACKGROUND

### A. Elastic energy functional

The surface density of energy of elastic deformations of lipid monolayer is written as [1]:

$$w = \frac{k_c}{2}\big[\text{div}(\mathbf{n}) + J_0\big]^2 - \frac{k_c}{2}J_0^2 + \frac{k_t}{2}\mathbf{t}^2 + \frac{k_a}{2}(\alpha - \alpha_0)^2 +$$
$$+ \frac{\sigma}{2}\big[\mathbf{grad}(H)\big]^2 + k_G K + \frac{k_{rot}}{2}\big[\mathbf{rot}(\mathbf{n})\big]^2,$$

(S1)

where $k_c$, $k_t$, $k_a$, $k_G$, $k_{rot}$ are elastic moduli of splay, tilt, lateral stretching, Gaussian splay, and twist, respectively; $J_0$ is the spontaneous curvature of the lipid monolayer; $\alpha_0 = \sigma/k_a$ is the spontaneous stretching caused by the constant lateral tension $\sigma$; in the Cartesian coordinate system, the $Oz$ axis of which is perpendicular to the membrane plane, $H = H(x, y)$ is the function determining the $z$-coordinates of points of the neutral surface. Let's denote the similar function for $z$-coordinates of points of the monolayer interface as $M = M(x, y)$. The surface density of energy of elastic deformations, Eq. (S1), is counted from the state of the planar unperturbed monolayer. The first term corresponds to the energy density of splay counted from the spontaneous state, which is characterized by spontaneous curvature $J_0$. The second term corresponds to the splay energy density of planar lipid monolayer counted from the spontaneous state. Thus, the difference between the first and the second terms is the splay energy density counted from the state of planar monolayer (zero curvature). The 5th term corresponds to the contribution of the lateral tension. This contribution is also counted from the state of the planar monolayer. The corresponding energy contribution should be written as: $\int \sigma dS - \sigma A_0$, where $A_0$ is the area of the monolayer surface in the planar unperturbed state, which is chosen as the reference state. This expression can be modified as: $\int \sigma dS - \sigma A_0 =$ $\int \sigma \sqrt{1 + \big[\mathbf{grad}(H)\big]^2}\,dxdy - \sigma A_0 \quad \approx \quad \int \sigma\left(1 + \frac{1}{2}\big[\mathbf{grad}(H)\big]^2\right)dxdy - \sigma A_0 \quad =$ $\sigma A_0 + \int \frac{\sigma}{2} \big[ \mathbf{grad}\,(H) \big]^2 dxdy - \sigma A_0 = \int \frac{\sigma}{2} \big[ \mathbf{grad}\,(H) \big]^2 dxdy$ . The term corresponding to the lateral tension in Eq. (S1) coincides with the expression under the integral.

The bulk modulus of lipid membranes is very large, $\sim 10^{10}$ J/m$^3$ [2]. The hydrophobic part of lipid monolayers can be considered volumetrically incompressible. The condition of volumetric incompressibility within the required accuracy reads [3]:

$$H - M = h - \frac{h^2}{2} \mathrm{div}\,(\mathbf{n}) - h\alpha, \tag{S2}$$

where $h$ is the thickness of the hydrophobic part of the lipid monolayer in the undeformed state. For simplicity, below we refer to $h$ as the monolayer thickness. This condition allows expressing $\alpha$ via $H$, $M$, and $\mathbf{n}$:

$$\alpha = \frac{1}{h} \bigg[ M + h - H - \frac{h^2}{2} \mathrm{div}\,(\mathbf{n}) \bigg]. \tag{S3}$$

For small deformations, $\mathbf{N} = \big( \partial H / \partial x, \partial H / \partial y, -1 \big)^T$, where $T$ is the transposition. Substituting $\mathbf{N}$ into the expression for the tilt vector $\mathbf{t} = \mathbf{n} - \mathbf{N}$, and the expressions for $K$, $\alpha$, $\mathbf{t}$ into the elastic energy density (S1), integrating over the neutral surface, one can obtain the elastic energy functional of the lipid monolayer:

$$W = \int dS \bigg\{ \frac{k_c}{2} \big[ \mathrm{div}\,(\mathbf{n}) + J_0 \big]^2 - \frac{k_c}{2} J_0^2 + \frac{k_t}{2} \big[ \mathbf{n} - \mathbf{grad}\,(H) \big]^2 +$$

$$+ \frac{\sigma}{2} \big[ \mathbf{grad}\,(H) \big]^2 + \frac{k_a}{2h^2} \bigg[ h - \frac{h^2}{2} \mathrm{div}\,(\mathbf{n}) + M - H - h\alpha_0 \bigg]^2 + \tag{S4}$$

$$+ k_G K + \frac{k_{rot}}{2} \big[ \mathbf{rot}\,(\mathbf{n}) \big]^2 \bigg\}.$$

A similar expression can be obtained for the elastic energy functional of the opposing lipid monolayer. These two functionals are not independent, as they include the common function $M$, characterizing the shape of the monolayer interface. The functional of the elastic energy of the lipid bilayer is given by the sum of the functionals of its constituent monolayers [4]. Below, all values related to the upper monolayer are designated by the index "$u$"; those for the lower monolayer — by the index "$l$".

## B. Hydrophobic defect

According to Marcelja's theory [5], the energy of the cylindrical hydrophobic cavity of the radius $R_0$ and the length $H_h$ coaxial with the $Oz$-axis filled with water can be obtained in the following form:

$$W_{hyd} = \sigma_0 \big( 2\pi R_0 H_h \big) \frac{I_1 \bigg( \dfrac{R_0}{\xi_h} \bigg)}{I_0 \bigg( \dfrac{R_0}{\xi_h} \bigg)}, \tag{S5}$$

where $I_0$, $I_1$ are corresponding modified Bessel functions; $\xi_h \approx 1$ nm is the characteristic length of hydrophobic interactions; $\sigma_0 \approx 40$ mN/m is the surface tension of the planar macroscopic water-lipid tails interface [3,6]; $(2\pi R_0 H_h)$ is the area of the side-surface of the cylinder. This energy is added to the energy of membrane deformations (S4). We assume that the hydrophobic cylinder of the hydrophobic defect meets the vertical monolayer neutral surface at coordinates $Z_{h1}$ and $Z_{h2}$. At these points, $R(Z_{h1}) = R(Z_{h2}) = R_0$. The following boundary condition should be imposed:

$$H_u\left(R_0\right)-H_d\left(R_0\right)=H_h=Z_{h2}-Z_{h1}.\tag{S6}$$

Based on the geometric meaning of the director, the projection of the director of the vertical monolayer onto the $Oz$-axis is assumed to have a jump at the hydrophobic cylinder:

$$v\left(Z_{h2}\right)-v\left(Z_{h1}\right)=-\frac{H_h}{\sqrt{\left(\dfrac{H_h}{2}\right)^2+\left(h-\dfrac{H_h}{2}\right)^2}}.\tag{S7}$$

### C. Boundary conditions

The functional of elastic energy should be supplemented by appropriate boundary conditions. In general, the monolayer director and the neutral surface are assumed to be continuous everywhere except for the region of the hydrophobic cylinder in the hydrophobic defect state. The deformations should be finite (and small) and should decay far from the pore and amphipathic peptides. In the non-local mechanism of pore formation, we impose only these general conditions along with the boundary conditions (S6), (S7) at the hydrophobic defect.

In the local mechanism of pre-pore formation, amphipathic peptides are explicitly considered as membrane-deforming inclusions. In the membrane region occupied by the inserted peptide, the neutral surface of the lipid monolayer does not exist, as the peptide pushes away polar lipid heads. We model the outline of the peptide at the neutral surface as a rectangle with rounded edges, the length of the smaller side of which is equal to the peptide diameter $2r_0$. The rounded edges are two half-circles at each smaller side of the rectangle; the radius of each half-circle is equal to the peptide radius $r_0$. The total length of the peptide in the monolayer plane (including the longer side of the rectangle and the two half-circles) is $L_p$ [Fig. S1(a)]. The boundary director is set at the outline of each peptide. The director has two components parallel to the membrane plane: normal to the outline, $\mathbf{n}_n$, and tangential to the outline, $\mathbf{n}_\tau$. We impose boundary conditions on the director components at the outline as follows:

$$\begin{aligned}\left|\mathbf{n}_n\left(\Gamma\right)\right|&\equiv n_0,\\ \left|\mathbf{n}_\tau\left(\Gamma\right)\right|&=0,\end{aligned}\tag{S8}$$

where $\Gamma$ is the boundary (outline) of the peptide at the neutral surface of the lipid monolayer. Using a simple geometric interpretation of the director, we estimate $n_0$ from the angle $\theta$ between the unit vector of the boundary director and the unit normal to the undeformed neutral surface of the monolayer, which yields $n_0\approx\sin\theta=r_0(r_0^2+h^2)^{-1/2}$ [Fig. S1(b)]. This estimation is rather qualitative, as the exact value of the normal component of the boundary director, $\mathbf{n}_n$, is determined by a complex interplay of the depth of insertion of the peptide into the lipid monolayer [Fig. S1(d), (e)], features of the chemical and physical interactions of the peptide with lipid polar heads and hydrophobic tails, etc. However, for melittin and magainin this estimation appears quite accurate [7]. The outlines of two peptides are allowed to rotate around the longitudinal axes of the peptides [Fig. S1(c)]. The rotation is characterized by the director $\mathbf{n_p}$, which is the average of the directors $\mathbf{n_1}$ and $\mathbf{n_2}$ at the right and left peptide boundaries, respectively. Upon the rotation, the total jump in coordinates of the neutral surface of the upper monolayer at the right and left boundaries of the peptide is $2\delta=2r_0n_{px}$, where $n_{px}$ is the projection of $\mathbf{n_p}$ onto the $Ox$-axis. The conditions (S8) imposed on the boundary director change accordingly to account for the peptide rotation.

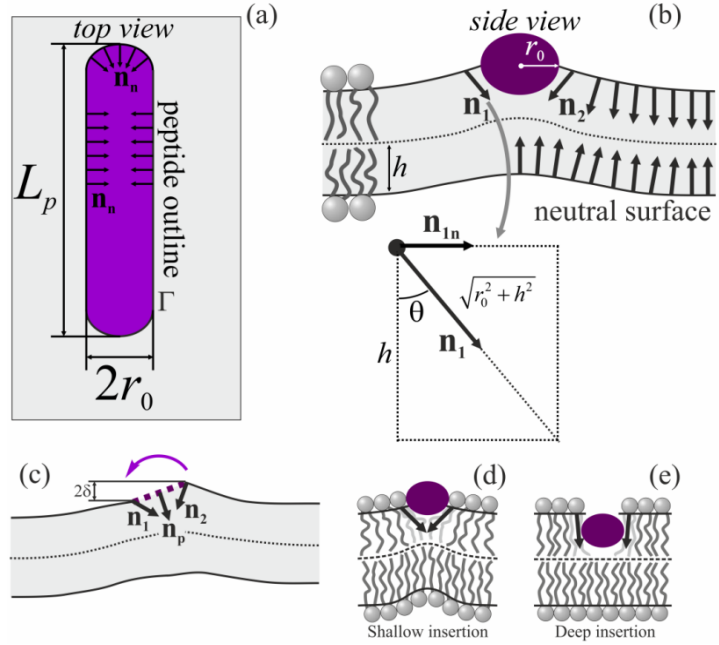

FIG. S1. Boundary conditions imposed by the amphipathic peptide. Panel (a) illustrates the outline of the peptide at the monolayer neutral surface; the outline is denoted $\Gamma$. The peptide diameter is $2r_0$, its length is $L_p$. Top view (i.e., down onto the membrane) is presented. In panel (b), the side view of the same peptide is presented. The director at the peptide boundary is set at the peptide outline $\Gamma$ at the upper monolayer neutral surface. Imposed boundary conditions fix the director component parallel to the neutral surface and normal to $\Gamma$ ($|\mathbf{n_n}(\Gamma)| = n_0$); the tangential component is set equal to zero ($|\mathbf{n_\tau}(\Gamma)| = 0$). The geometric interpretation of the boundary director is illustrated below. Peptides are allowed to rotate around their longitudinal axes, as illustrated in panel (c). The non-zero average director $\mathbf{n_p} = (\mathbf{n_1} + \mathbf{n_2})/2$ characterizes the rotation. Upon the rotation, the left and right boundaries of the peptide shift upward or downward, respectively; the total jump in coordinates of the neutral surface at the right and left boundaries of the peptide is $2\delta = 2r_0 n_{px}$ ($n_{px}$ is the projection of $\mathbf{n_p}$ onto the $Ox$-axis). The value of the boundary director should depend on the depth of the peptide insertion; generally, for shallowly inserted peptide [panel (d)] the absolute value of the boundary director (half of the angle between two thick black arrows) should be larger than for deeply inserted peptide [panel (e)]. The boundary director determines a curvature induced by the inserted peptide [compare the shapes of the membrane in panels (d) and (e)].

## E. Rotationally symmetric system

In the case when the system possesses the rotational symmetry, the functional of the elastic energy (S4) can be minimized analytically. We introduce a cylindrical coordinate system $Orz$ whose $Oz$ axis coincides with the axis of the rotational symmetry. We place the origin of the coordinate system $O$ onto the plane of the monolayer interface. In the rotationally symmetric system, all vector quantities can be replaced with their projections onto the radial axis $Or$: $\mathbf{n} \to n_r \equiv n$, $\mathbf{t} \to t_r \equiv t$, $\mathbf{N} \to N_r \equiv N$. In addition, within the required accuracy, $\mathbf{rot}(\mathbf{n}) = \mathbf{0}$, $\mathrm{div}(\mathbf{n}) \to n' + n/r$, $\mathbf{grad}(H) \to H'$, where the prime in the superscript denotes the derivative with respect to the coordinate $r$. The elastic energy functional (S4) can be rewritten for the bilayer in the general case of different lateral tensions in the upper and lower monolayers as follows:

$$W = \int 2\pi r \frac{k_t}{2} \left\{ l^2 \left( n_u' + \frac{n_u}{r} + J_0 \right)^2 - l^2 J_0^2 + \left( n_u - H_u' \right)^2 + \right.$$

$$+ A \left( 1 - \frac{H_u - M}{h} - \frac{h}{2} \left( n_u' + \frac{n_u}{r} \right) - \alpha_{0u} \right)^2 + 2 s_u \left( H_u' \right)^2 + 2 \frac{k_G}{k_t} n_u' \frac{n_u}{r} \right\} dr +$$

$$+ \int 2\pi r \frac{k_t}{2} \left\{ l^2 \left( n_l' + \frac{n_l}{r} + J_0 \right)^2 - l^2 J_0^2 + \left( n_l + H_l' \right)^2 + \right. \tag{S9}$$

$$+ A \left( 1 - \frac{M - H_l}{h} - \frac{h}{2} \left( n_l' + \frac{n_l}{r} \right) - \alpha_{0l} \right)^2 + 2 s_l \left( H_l' \right)^2 + 2 \frac{k_G}{k_t} n_l' \frac{n_l}{r} \right\} dr,$$

where $l^2 = k_c/k_t$; $A = k_a/k_t$; $s_u = \sigma_u/(2k_t)$; $s_l = \sigma_l/(2k_t)$. Here the conditions of local volumetric incompressibility (S3) were utilized. Variation of this functional with respect to the functions $n_u(r)$, $n_l(r)$, $H_u(r)$, $H_l(r)$, $M(r)$ yields five Euler-Lagrange equations:

$$\left( l^2 + \frac{Ah^2}{4} \right) \left( n_u'' + \frac{n_u'}{r} \right) - \frac{1}{r^2} \left( r^2 + l^2 + \frac{Ah^2}{4} \right) n_u + \left( 1 + \frac{A}{2} \right) H_u' - \frac{A}{2} M' = 0,$$

$$\left( l^2 + \frac{Ah^2}{4} \right) \left( n_l'' + \frac{n_l'}{r} \right) - \frac{1}{r^2} \left( r^2 + l^2 + \frac{Ah^2}{4} \right) n_u - \left( 1 + \frac{A}{2} \right) H_l' + \frac{A}{2} M' = 0,$$

$$\left( 1 + \frac{A}{2} \right) \left( n_u' + \frac{n_u}{r} \right) - \left( 2 s_u + 1 \right) \left( H_u'' + \frac{H_u'}{r} \right) + \frac{A}{h^2} \left( H_u - h - M + h \alpha_{0u} \right) = 0, \tag{S10}$$

$$\left( 1 + \frac{A}{2} \right) \left( n_l' + \frac{n_l}{r} \right) + \left( 2 s_l + 1 \right) \left( H_l'' + \frac{H_l'}{r} \right) - \frac{A}{h^2} \left( H_l + h - M - h \alpha_{0l} \right) = 0,$$

$$n_u' - n_l' + \frac{n_u - n_l}{r} + \frac{2}{h^2} \left( H_u + H_l - 2M + h \left( \alpha_{0u} - \alpha_{0l} \right) \right) = 0.$$

It is noteworthy that neither $k_G$ nor $J_0$ stand in equations (S10). This is because the corresponding terms in the functional (S9) can be explicitly integrated thus reducing to constants that depend on the values of the director at the boundary of the integration domain. Indeed, the terms comprising $J_0$ and $k_G$ can be transformed as follows:

$$W_0 + W_G = \int_{R_1}^{R_2} 2\pi r \frac{k_t}{2} l^2 J_0 \left( n' + \frac{n}{r} \right) dr + \int_{R_1}^{R_2} 2\pi r k_G n' \frac{n}{r} dr =$$

$$= \pi k_t l^2 J_0 \int_{R_1}^{R_2} \left( n' r + n \right) dr + 2\pi k_G \int_{R_1}^{R_2} \left( n' n \right) dr = \pi k_t l^2 J_0 \int_{R_1}^{R_2} \left( n r \right)' dr + \pi k_G \int_{R_1}^{R_2} \left( n^2 \right)' dr = \tag{S11}$$

$$= \pi k_t l^2 J_0 \left( n r \right) \Big|_{R_1}^{R_2} + \pi k_G \left( n^2 \right) \Big|_{R_1}^{R_2} = \pi k_t l^2 J_0 \left( n \left( R_2 \right) R_2 - n \left( R_1 \right) R_1 \right) + \pi k_G \left( n^2 \left( R_2 \right) - n^2 \left( R_1 \right) \right).$$

Let's enumerate the equations in (S10) from $E_1$ to $E_5$. From the last equation, $E_5$, one can express $M(r)$:

$$M = \frac{h^2}{4} \left( n_u' - n_l' + \frac{n_u - n_l}{r} + \frac{2}{h^2} \left( H_u + H_l + h \left( \alpha_{0u} - \alpha_{0l} \right) \right) \right), \tag{S12}$$

and substitute it into the first four equations $E_1, \ldots, E_4$ of the system (S10). Further, we introduce new functions: $s = n_u + n_l$, $d = n_u - n_l$. The combination $(E_1 + E_2)/(1 + A/2) + (E_1 - E_2)$ allows to express $H_u'$ through $s$, $d$ and their derivatives:

$$H_u' = -\frac{Ah^2 + 4l^2}{4(2 + A)} \left( s'' + \frac{s'}{r} \right) - \frac{l^2}{2} \left( d'' + \frac{d'}{r} \right) + \frac{4r^2 + Ah^2 + 4l^2}{4(2 + A) r^2} s + \frac{l^2 + r^2}{2r^2} d. \tag{S13}$$

The combination $(E_1 + E_2)/(1 + A/2) - (E_1 - E_2)$ allows to get $H_l'$:

$$H_l' = \frac{Ah^2 + 4l^2}{4(2+A)}\left(s'' + \frac{s'}{r}\right) - \frac{l^2}{2}\left(d'' + \frac{d'}{r}\right) - \frac{4r^2 + Ah^2 + 4l^2}{4(2+A)r^2}s + \frac{l^2 + r^2}{2r^2}d. \quad (S14)$$

Substitution of $M(r)$ from Eq. (12), $H_u'$ from Eq. (S13) and $H_l'$ from Eq. (14) into equations $E_1$ and $E_2$ of the system (10) yields two equations for the functions $s(r)$ and $d(r)$. Further, by utilizing standard transformations, these two equations can be reduced to a single linear differential equation of the eighth order for the function $d(r)$. The equation is quite bulky, and thus we did not present it here. By direct substitution, we obtained that the function $c_0/r$ ($c_0$ is some constant coefficient) is the solution of this equation. In addition, the substitution of the functions $d(r) = J_1(\lambda r)$ or $d(r) = Y_1(\lambda r)$ ($J_1$, $Y_1$ are the corresponding Bessel functions of the first order) yields the following characteristic polynomial equation:

$$\lambda^2\left\{\lambda^6 h^2 l^2\left(4l^2 + Ah^2\right)\left(1 + 2\left(s_u + s_l\right) + 4s_u s_l\right) + \right.$$
$$+\lambda^4\left(\left(h^2 + 2l^2\right)^2\left(s_u + s_l\right) + 4h^2\left(8l^2 + h^2\right)s_u s_l - 4Al^2\left(h^2 - l^2\right)\right) + \quad (S15)$$
$$\left. + \lambda^2\left(16h^2 s_u s_l + 4A\left(2l^2 - h^2\right)\left(s_u + s_l\right) + 4Al^2\right) + 4A\left(s_u + s_l\right)\right\} = 0.$$

This equation has two trivial roots $\lambda = 0$, and three pairs of non-trivial roots: $\pm p_1$, $\pm p_2$, $\pm p_3$; $p_1$ and $p_2$ are complex-conjugate, while $p_3$ is imaginary for typical values of the membrane elastic parameters. All $p_i$ are different for common lipid parameters. The expression for $d(r)$ can be written as follows:

$$d(r) = c_0\frac{1}{r} + c_1 r + c_2 J_1\left(p_1 r\right) + c_3 J_1\left(p_2 r\right) + c_4 J_1\left(p_3 r\right) + c_5 Y_1\left(p_1 r\right) + c_6 Y_1\left(p_2 r\right) + c_7 Y_1\left(p_3 r\right), \quad (S16)$$

where $c_0$, $c_1$, …, $c_7$ are constant complex coefficients that should be determined from the boundary conditions. The successive substitution of $d(r)$ into intermediate equations allows to obtain explicitly all the functions $n_u(r)$, $n_l(r)$, $H_u(r)$, $H_l(r)$, $M(r)$. However, the corresponding expressions are very bulky; thus, we did not present them here. Further, the functions $n_u(r)$, $n_l(r)$, $H_u(r)$, $H_l(r)$, $M(r)$ were substituted into the energy functional (S9); after the integration, we obtained analytically the elastic energy of the deformed lipid bilayer. In the case of symmetric lateral tension, $s_u = s_l = s_0$, the inverse characteristic lengths of the deformations are given by:

$$p_{1,2} = \frac{2}{h}\sqrt{\frac{A\left(h^2 - l^2\right) - h^2 s_0 \pm \sqrt{h^4 s_0 - A\left(2h^2\left(l^2 + h^2\right) + 4l^2 h^2\right) - A^2\left(h^4 s_0 + l^2\left(2h^2 - l^2\right)\right)}}{2\left(1 + s_0\right)\left(4l^2 + Ah^2\right)}},$$

$$p_3 = \frac{1}{l}\sqrt{\frac{-s_0}{1 + s_0}}. \quad (S17)$$

The symbolic computations are available upon request.

## F. Vertical monolayer region

The energy functional (S9) was derived under the assumption of small deformations. Formally, the following inequalities should hold: $|n| \ll 1$, $|n'h| \ll 1$, meaning that the director should deviate but slightly from the unit normal vector to the surface of the flat undeformed bilayer, i.e., in our coordinate system the director should be almost parallel to the $Oz$-axis. However, in the equatorial plane of the pore, the director is almost perpendicular to the $Oz$-axis, i.e., $|n| \approx 1$. Thus, the energy functional (S9) is inapplicable in the vicinity of the pore equator. However, the deformations at the pore edge can still be considered small if a vertical cylindrical monolayer coaxial with the $Oz$-axis is considered as the reference state instead of the planar horizontal bilayer. This consideration allows to formally apply the linear theory of elasticity.

The zone of the vertical cylindrical monolayer is conjugated with the bilayer region of the membrane based on the continuity of neutral surfaces and directors. The shape of the neutral surface of the vertical monolayer is described by the $r$-coordinates of its points, $R(z)$, the average orientation of the lipid molecules — by the projection of the director onto the $Oz$-axis, $v(z)$. Within the required accuracy, $\text{div}(\mathbf{n}) = dv(z)/dz + 1/R(z)$. The neutral surface and the monolayer interface are assumed to deviate slightly from cylindrical surfaces of radii $R_v$ and $M_v$, respectively, i.e., $R(z) = R_v + u(z)$, $M(z) = M_v + m(z)$, where $|u(z)| \ll R_v$, $|m(z)| \ll M_v$. The functions $R(z)$ and $M(z)$, as well as the constants $R_v$ and $M_v$, should satisfy the conditions of local volumetric incompressibility:

$$M(z) - R(z) = h - \frac{h^2}{2}\left(\frac{dv(z)}{dz} + \frac{1}{R(z)}\right) - h\beta(z),$$

$$M_v - R_v = h - \frac{h^2}{2R_v^2}, \tag{S18}$$

where $\beta(z)$ denotes the relative lateral stretching of the neutral surface of the vertical cylindrical monolayer. From these equations the deviation of the neutral surface from the cylindrical one, $u(z)$, can be expressed in the linear order of smallness as follows:

$$u(z) = \frac{R_v^2}{2R_v^2 + h^2}\left(2m(z) + 2h\beta(z) + h^2\frac{dv(z)}{dz}\right). \tag{S19}$$

The functional of the elastic energy of the vertical monolayer can thus be written as:

$$W_m = \frac{k_t}{2}\int_{Z_1}^{Z_2} 2\pi R\sqrt{1 + \left(\frac{dR}{dz}\right)^2}\left\{l^2\left(\frac{dv}{dz} + \frac{1}{R} + J_0\right)^2 - \right.$$

$$\left. -l^2 J_0^2 + (v - N_z)^2 + A(\beta - \beta_0)^2 + s\right\}dz + 2\pi k_G\left(v(Z_2) - v(Z_1)\right). \tag{S20}$$

Here $s = \sigma/k_t$, $\beta_0 = \sigma/k_a$. Within the required accuracy, $N_z = -du(z)/dz$; $u(z)$ is defined by Eq. (S19). We took Taylor series of the functional (S20) up to the second order with respect to small functions $v(z)$, $m(z)$, $\beta(z)$ and their derivatives. Further, we took the first variation of the resulting functional with respect to functions $v(z)$, $m(z)$, $\beta(z)$. This led to three Euler-Lagrange linear differential equations, whose general solution could be written as:

$$\beta(z) = \beta_0,$$

$$v(z) = d_1 e^{-o_1 z} + d_2 e^{o_1 z} + d_3 e^{-o_2 z} + d_4 e^{o_2 z},$$

$$m(z) = \frac{\left(2R_v^2 + h^2\right)\left(l^2 - 2\sigma R_v^2\right)}{4R_v l^2} +$$

$$+ D_1 d_1 e^{-o_1 z} + D_2 d_2 e^{o_1 z} + D_3 d_3 e^{-o_2 z} + D_4 d_4 e^{o_2 z}, \tag{S21}$$

where $d_1$, $d_2$, $d_3$, $d_4$ are complex constant coefficients, which should be determined from the boundary conditions; $D_1$, $D_2$, $D_3$, $D_4$ are known combinations of elastic parameters of the membrane that are too bulky to be presented here;

$$o_{1,2} = 2\left(l^2 + 2l^2 J_0 R_v + 2s R_v^2 + 2R_v^2\right)^{-\frac{1}{2}} \times$$

$$\times\left[2\sigma R_v^4 + 6l^2 J_0 R_v^3 - 2l^4 J_0^2 R_v^2 + l^2 R_v^2 + 2l^4 \pm \right.$$

$$\pm\left(2\sigma R_v^4 - 2l^4 + 5l^2 R_v^2 + 6l^2 J_0 R_v^3 - 4l^4 J_0 R_v^4 - 2l^4 J_0^2 R_v^2\right)^{\frac{1}{2}} \times$$

$$\left. \times\left(2\sigma R_v^4 - 2l^4 - 3l^2 R_v^2 + 6l^2 J_0 R_v^3 + 4l^4 J_0 R_v^4 - 2l^4 J_0^2 R_v^2\right)^{\frac{1}{2}}\right]^{\frac{1}{2}} \tag{S22}$$

are inversed characteristic lengths of deformations of the vertical monolayer region.

The vertical monolayer is conjugated with the upper monolayer of the bilayer part of the membrane along the circle $\{R_u, Z_u\}$; with the bottom monolayer — along the circle $\{R_l, Z_l\}$.

Within the required accuracy, the conjugation conditions following from the continuity of neutral surfaces and directors can be written as:

$$H_u\left(R_u\right)=Z_u, \quad R\left(Z_u\right)=R_u, \quad n_u\left(R_u\right)-v\left(Z_u\right)=1,$$
$$H_l\left(R_l\right)=Z_l, \quad R\left(Z_l\right)=R_l, \quad n_l\left(R_l\right)+v\left(Z_l\right)=1. \tag{S23}$$

We wrote the continuity equations (S23) in a qualitative, linear form, which greatly simplified the solution of the minimization problem. The above-described boundary conditions allowed to determine some of the unknown constant coefficients appearing in the general solutions of Euler-Lagrange equations (S10). The remaining unknown coefficients were analytically determined from the condition of the total energy minimum. The total energy was numerically minimized with respect to the coordinates of conjugation of the horizontal bilayer and vertical monolayer parts of the membrane, as well as with respect to the coordinates of the ends of the hydrophobic cylinder $Z_{h1}$, $Z_{h2}$ using the gradient descent method [8]. The symbolic computations and optimization programs are available upon request.

**RESULTS**

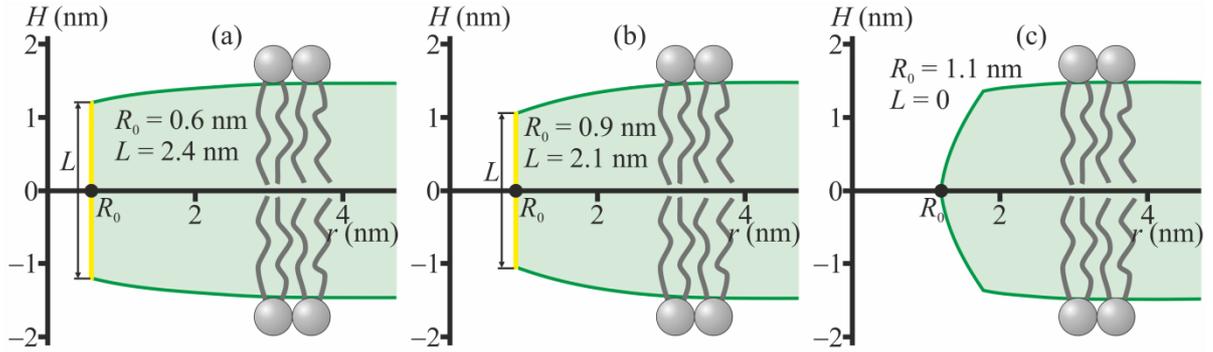

Figure S2. Membrane shape in the vicinity of the pore edge in the membrane with equal lateral tensions in both monolayers, with no peptides. The membrane shape is mirror symmetric with respect to the planar monolayer interface, as illustrated for $\sigma_u = \sigma_l = 2.5$ mN/m and (a) $R_0 = 0.6$ nm, $L = 2.4$ nm; (b) $R_0 = 0.9$ nm, $L = 2.1$ nm; (c) $R_0 = 1.1$ nm, $L = 0$. The yellow vertical line corresponds to the side-wall of the hydrophobic cylinder.

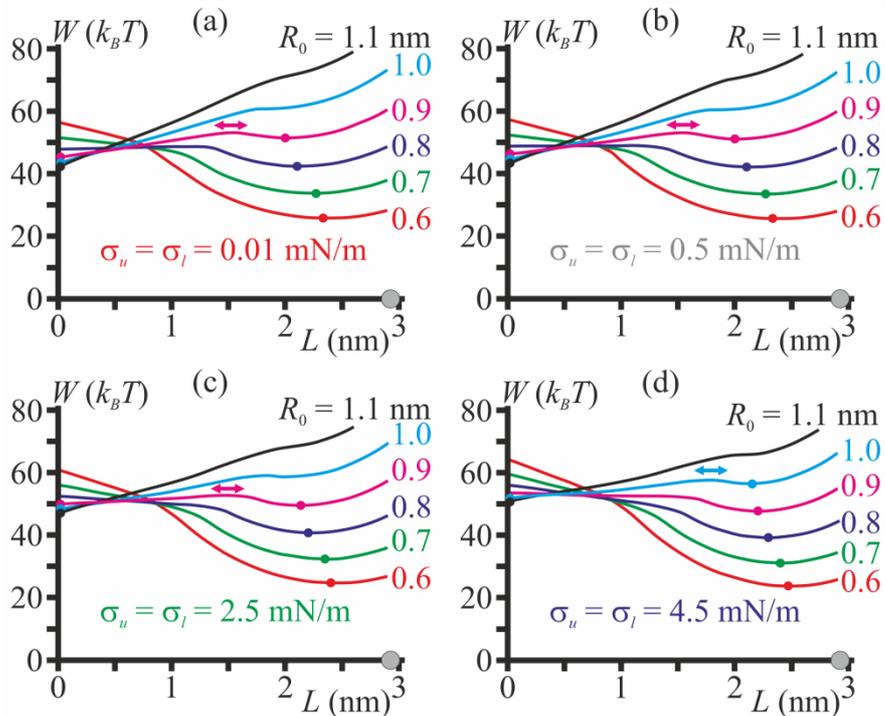

Figure S3. Pore energy landscapes in the membrane whose monolayers are subjected to equal lateral tensions with no peptides. Energy landscapes $W(R_0, L)$ for: (a) $\sigma_u = \sigma_l = 0.01$ mN/m; (b) $\sigma_u = \sigma_l = 0.5$ mN/m; (c) $\sigma_u = \sigma_l = 2.5$ mN/m; (d) $\sigma_u = \sigma_l = 4.5$ mN/m. $R_0$ is the parameter of the curves; its values (in nanometers) are indicated at each curve. Energy minima with respect to $L$ are shown as circles at each curve. $L = 2h = 2.9$ nm corresponds to the planar undisturbed bilayer. This state is indicated by the gray circle; the energy of such configuration is zero.

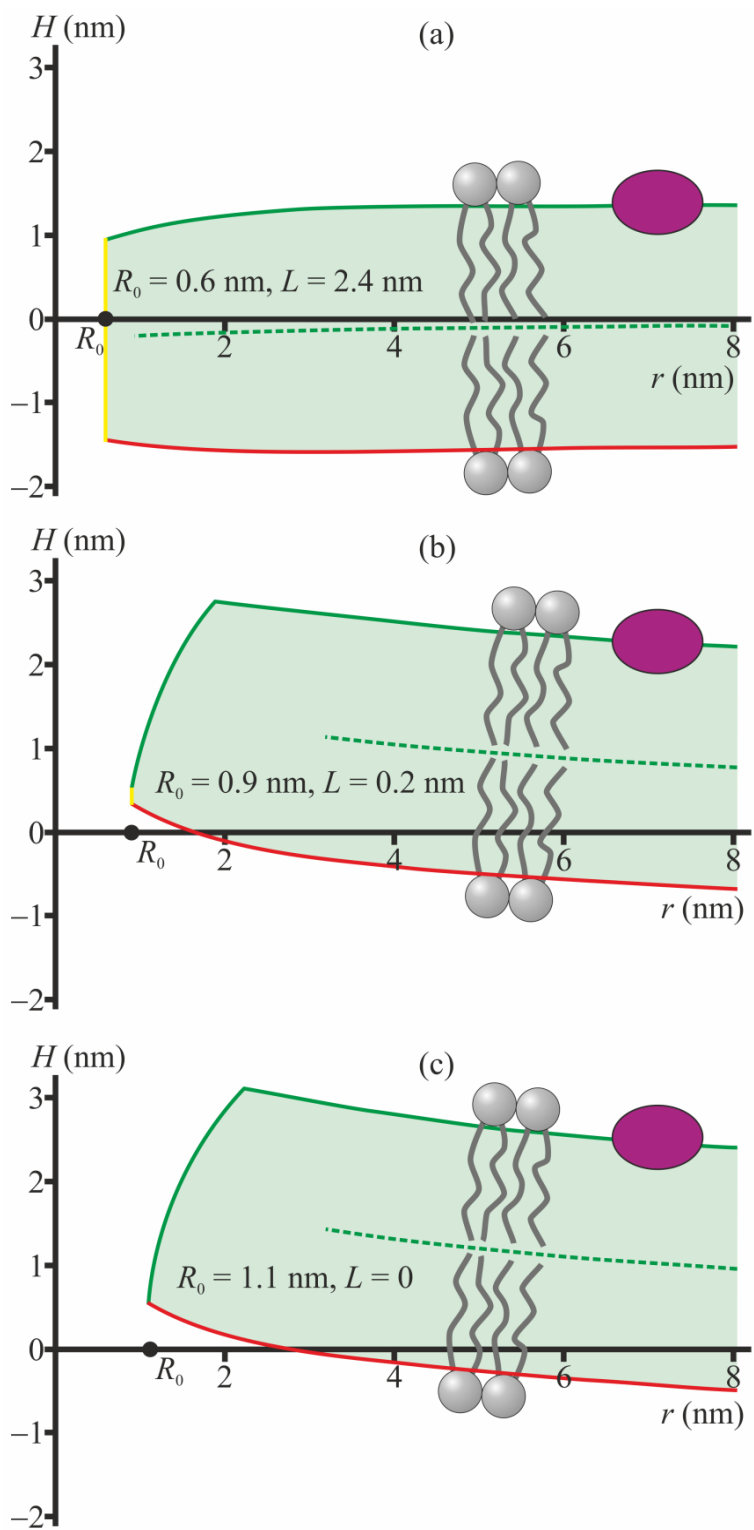

Figure S4. Shape of the membrane with the upper monolayer subjected to the lateral pressure ($\sigma_u$ = −2.5 mN/m) and the lower one — to the lateral tension ($\sigma_l$ = 2.5 mN/m). (a) Hydrophobic defect, $R_0$ = 0.6 nm, $L$ = 2.4 nm; (b) hydrophobic defect, $R_0$ = 0.9 nm, $L$ = 0.2 nm; (c) hydrophilic pore, $R_0$ = 1.1 nm, $L$ = 0. The yellow vertical line corresponds to the side-wall of the hydrophobic cylinder. Amphipathic peptides incorporated into the upper monolayer are schematically shown as violet ellipses.

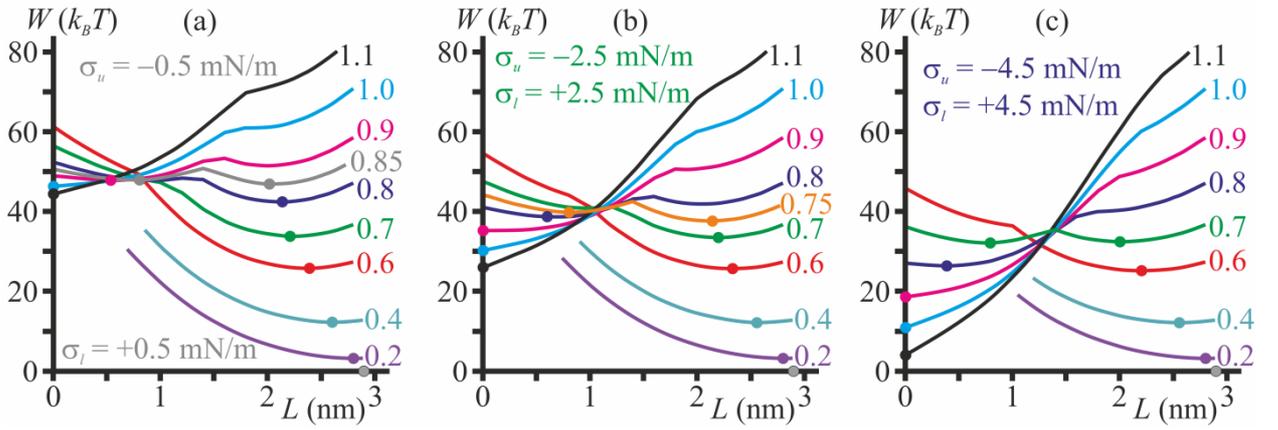

Figure S5. Pore energy landscapes in the membrane with the upper monolayer subjected to the lateral pressure and the lower one — to the lateral tension. The energy landscapes $W(R_0, L)$ for: (a) $\sigma_u = -0.5$ mN/m $+ \sigma_e$, $\sigma_l = 0.5$ mN/m $+ \sigma_e$; (b) $\sigma_u = -2.5$ mN/m $+ \sigma_e$, $\sigma_l = 2.5$ mN/m $+ \sigma_e$; (c) $\sigma_u = -4.5$ mN/m $+ \sigma_e$, $\sigma_l = 4.5$ mN/m $+ \sigma_e$. $R_0$ is the parameter of the curves; its values (in nanometers) are indicated at each curve. The energy minima with respect to $L$ are shown as circles at each curve. $L = 2h = 2.9$ nm corresponds to a planar undisturbed bilayer. This state is indicated by the gray circle; the elastic energy of such configuration is zero.

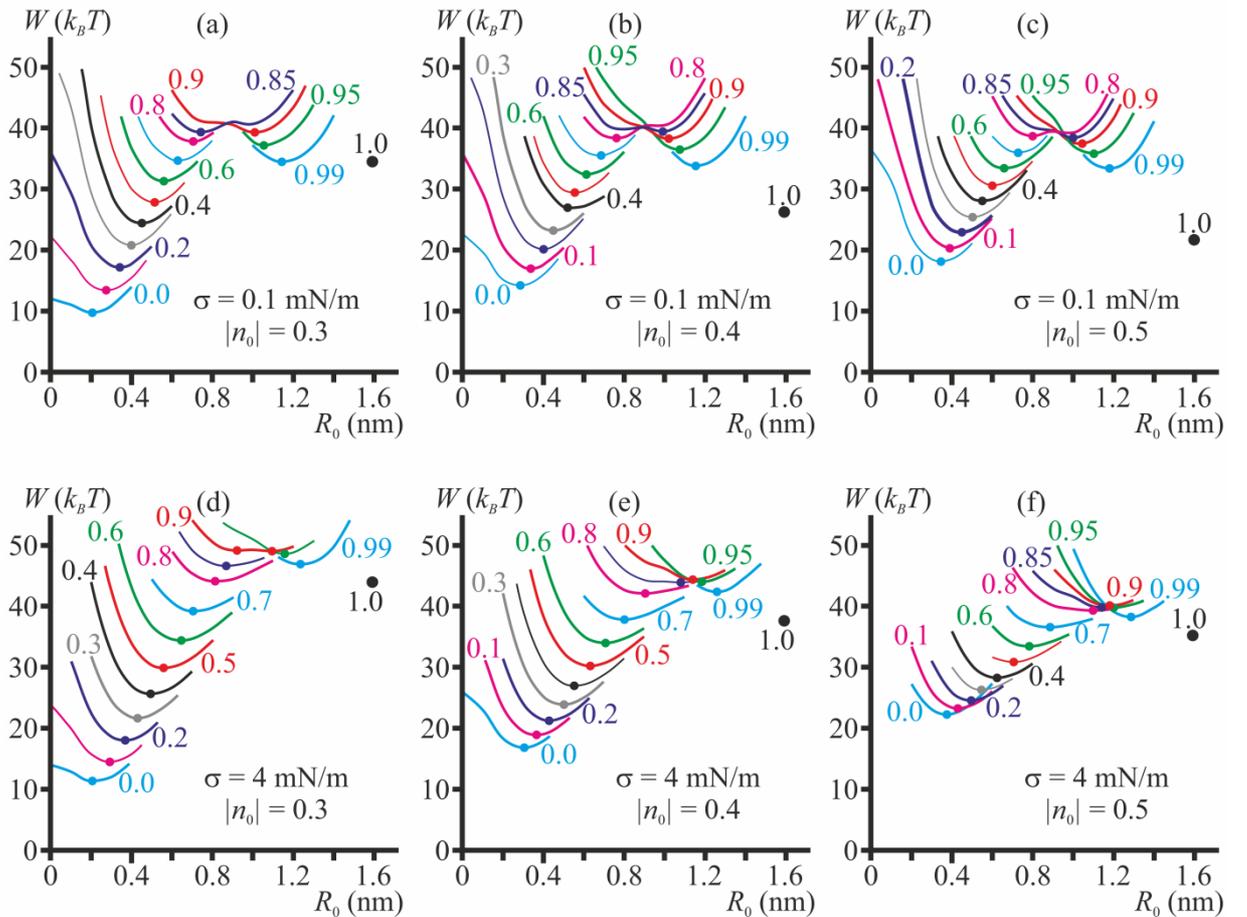

Figure S6. The energy surface $W(n_t, R_0)$ shown for fixed discrete values of $n_t$ and variable $R_0$ for a low lateral tension [$\sigma = 0.1$ mN/m per monolayer, panels (a)–(c)] and a high lateral tension [$\sigma = 4$ mN/m per monolayer, panels (d)–(f)] and different values of the boundary director projections [$|n_0| = 0.3$ in panels (a), (d); $|n_0| = 0.4$ in panels (b), (e); $|n_0| = 0.5$ in panels (c), (f)]. The dependencies $W(R_0)$ at each fixed value of $n_t$ have global minima, denoted by filled circles. These minima determine the optimal trajectories $R_0(n_t)$ of the pore formation. The highest

minimum in each panel corresponds to the saddle-point of the energy surface $W(n_t, R_0)$, which corresponds to the top of the energy barrier of the pore formation process.